\documentclass[a4paper,12pt]{report}
\usepackage{geometry}
\usepackage{graphicx}
\usepackage{amsmath}
\usepackage{amssymb}
\usepackage[breaklinks=true]{hyperref}

\usepackage{titling}

\pretitle{\begin{center}\Huge\bfseries}
\posttitle{\par\end{center}\vskip 0.5em}
\preauthor{\begin{center}\Large\ttfamily}
\postauthor{\end{center}}
\predate{\par\large\centering}
\postdate{\par}

\title{Relaxation in scalar gravitational field theory}
\author{Riccardo Fantoni}
\date{\today}

\begin{document}           

\maketitle
\thispagestyle{empty}
\begin{abstract}
We revisit the problem of relaxation in scalar gravitational field theory proposing a 
novel numerical solution to the problem.
\end{abstract}

\newcommand{\bq}{\begin{eqnarray}}
\newcommand{\eq}{\end{eqnarray}}
\newcommand{\utlr}{\tilde{u}_r}
\newcommand{\utlp}{\tilde{u}_\phi}
\newcommand{\utt}{{\tilde{u}^0}}
\newcommand{\rp}{r_{\!\! \scriptscriptstyle p}}
\newcommand{\rqs}{r_{\!\! \scriptscriptstyle q.s.}}
\newcommand{\xqs}{\xi_{\scriptscriptstyle q.s.}}
\newcommand{\aqs}{a_{\scriptscriptstyle q.s.}}
\newcommand{\xp}{{x^\prime}}
\newcommand{\xb}{{\bf x}}
\newcommand{\zb}{{\bf z}}
\newcommand{\xbp}{{\bf x^\prime}}
\newcommand{\tp}{{t^\prime}}
\newcommand{\rpp}{{r^\prime}}
\newcommand{\tep}{{\theta^\prime}}
\newcommand{\shalf}{{\scriptstyle \frac{1}{2}}}
\newcommand{\ssixth}{{\scriptstyle \frac{1}{6}}}
\newcommand{\rr}{{\bf r}}

\tableofcontents 

\setcounter{page}{1}
\chapter[Motivations]{Motivations}

In this report we want to predict the approach to equilibrium of a 
spherically symmetric field-particle system initially excited in a 
non-equilibrium state where the particle is in an unstable circular 
orbit around the origin \cite{Shapiro92}.

In particular we will be concerned with the realization
of a quasistatic approximation to the exact dynamical problem. As the
newly built gravitational wave detectors are preparing to receive
their first set of data, theorical efforts are being carried on to
solve exactly Einstein's equation to be able to timely interpret such data. 
Our quasistatic approximation in an unstable circular orbit could 
become an important tool in the event that such theoretical efforts fail 
to solve the exact problem in time.  
The approximation should be particularly useful in interpreting the
waveform coming from slowly decaying binary neutron stars. 

Binary neutron stars are known to exist and for some of the systems in
our own galaxy (like the relativistic binary radio pulsar PSR B1913+16
and PSR B1534+12) general relativistic effects in the binary orbit
have been measured to high precision. With the construction of laser
interferometers well underway, it is of growing urgency that we be
able to predict theoretically the gravitational waveform emitted
during the inspiral and the final coalescence of the two
stars. Relativistic binary systems, like binary neutron stars and
binary black holes pose a fundamental challenge to theorists, as the
two-body problem is one of the outstanding unsolved problems in
classical general relativity.

\chapter[Introduction]{Introduction}
\label{intro}

When studying a two body problem one decomposes it in the 
trivial problem involving the center of mass motion and 
the harder one involving the relative motion of the two 
masses. Is the second one, we want to focus on.
Since we don' t want to deal with all the difficulties of 
General Relativity (there is no analytic solution to the two body 
problem in GR) and we want to have a more realistic theory than the 
Newtonian one, we choose to employ a theory which describes gravitation 
by a nonlinear scalar gravitational field $\Phi$ in special relativity.
To decribe the relative motion in a two body problem
we just need one particle moving around the origin. The particle motion 
is confined at all times in its orbital plane, and its position there 
is determined by the distance from the origin $r_p$, and the azimuthal 
angle $\phi_p$. To follow the dynamical evolution of the field-particle 
system in scalar gravity, one needs to solve a single hyperbolic partial
differential equation describing the field evolution, coupled to a system 
of two ordinary differential equations describing the particle motion,
\bq \label{problem}
\boxed{\left\{
\begin{array}{l}
\square \Phi(\rr,t)=\mbox{source}~~,\\
\ddot{r}_p=\ldots~~,\\
\ddot{\phi}_p=\ldots~~.
\end{array}
\right.}
\eq
The source term of the field equation is where the coupling between 
the field and the particle dynamics takes place, and is responsible for 
the nonlinearity of the problem:
source $\sim\exp(\Phi)\rho$, where $\rho(\rr,t)=(m/\gamma)\delta(\rr-
\rr_p(t))$ is the comoving matter density, $m$ the particle rest 
mass, and $\gamma$ the Lorentz factor.

In particular we want to study the even simpler, spherically symmetric 
problem. It is infact a peculiarity of scalar gravitation that of being 
able to generate gravitational waves even in spherical symmetry. 
This allows the study of wave generation and propagation with the use of 
just one spatial dimension plus time.
In spherical symmetry the particle angular momentum $\utlp$ is conserved.
There are then three important quantities in our problem: the initial 
distance from the origin $r_i$, the particle rest mass $m$, and its
angular momentum $\utlp$. Two adimensional combinations of these quantities
are particularly important to parametrize the problem: 
\begin{description}
\item[(1)] The initial compaction $\alpha=r_i/m$ which tunes the 
nonlinearity of the problem:
\begin{description}
\item[$\alpha \gg 1$] The system is in a weak field and slow particle
velocity regime. Newtonian gravitation provides a good analytical
approximation to the nearly linear and periodic system.
\item[$\alpha \sim 1$] The system is nonlinear and aperiodic. There
is no analytic solution to the coupled equations (\ref{problem}), 
and a numerical integration is needed. In this report we will
describe an approximate solution which works well when the system relaxes 
slowly.
\end{description}
\item[(2)] An adimensional measure of the particle angular momentum 
$J=\utlp/(\utlp)_{circ}(r_i)$. Here we are indicating with 
$(\utlp)_{circ}(r_i)$ the angular momentum that the particle should 
have in order to be in a circular orbit at the initial radius $r_i$.
\begin{description}
\item[$J=0$] The particle collapses to the origin.
\item[$J=1$] The particle is in a stable circular orbit. Even though 
the particle is in circular motion around the origin, it doesn' t loose
energy by gravitational radiation because in spherical symmetry the 
particle in circular orbit represents a stationary spherical mass shell. 
\item[$J >1$] The particle is initially at the periastron of its 
elliptical orbit. There is a value $J_e$  such that if $J >J_e$ 
the particle escapes to infinity, if $J <J_e$ the particle orbit 
becomes circular at $t=\infty$, and of radius $r_e$ bigger than $r_i$. 
\begin{figure}[hbt]
\begin{center}
\includegraphics[width=10cm]{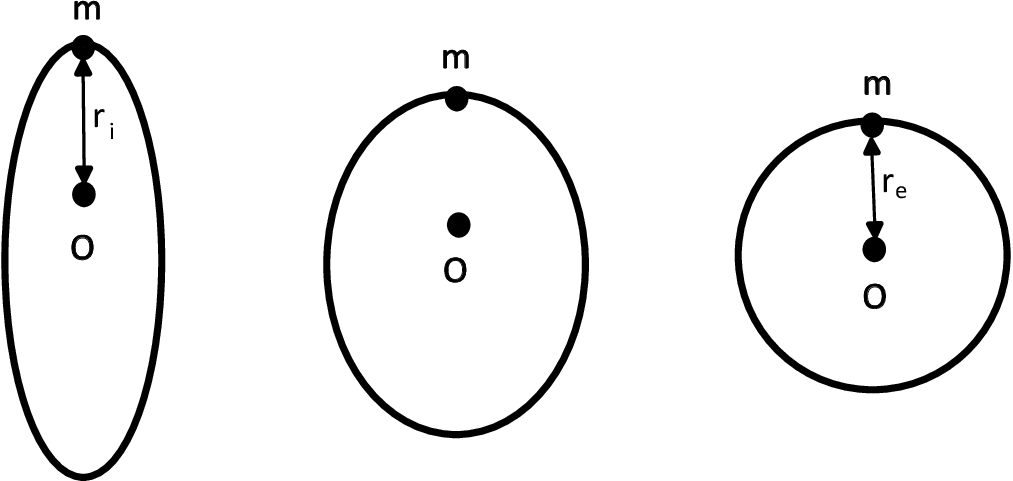}\\
\end{center}
\caption{Pictorial evolution of the particle orbit when $J>1$.
\label{J>1}}
\end{figure}
\item[$J <1$] The particle is initially at the apastron of its 
elliptical orbit. The particle orbit becomes circular at $t=\infty$, 
and of radius $r_e$ smaller than $r_i$. If $J \ll 1$
the shell relaxation will be fast (it will reach $r_e$ in a small 
number of oscillations) and the quasistatic approximation that we are 
now going to describe will break down.
\begin{figure}[hbt]
\begin{center}
\includegraphics[width=10cm]{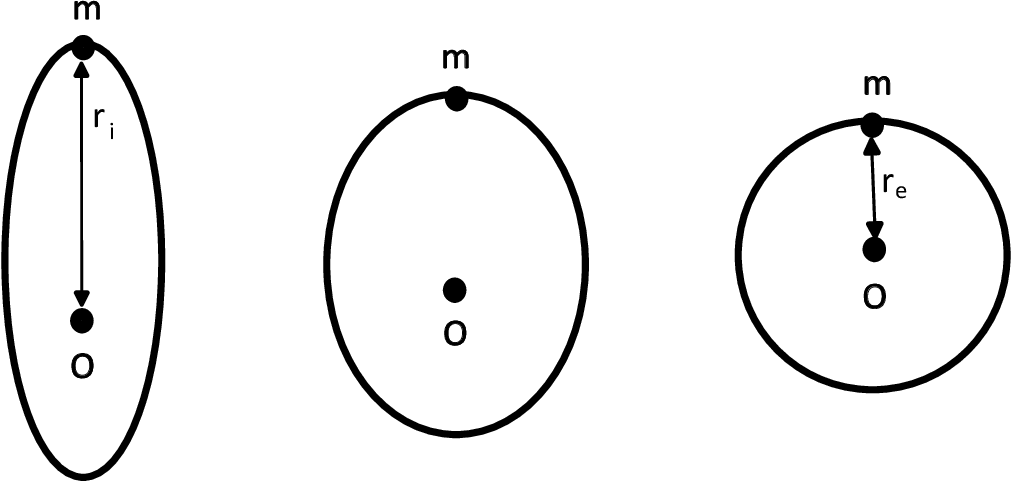}\\
\end{center}
\caption{Pictorial evolution of the particle orbit when $J<1$.
\label{J<1}}
\end{figure}
\end{description}
\end{description}

When the timescale of orbital decay by radiation is much longer than the
orbital period, the particle can be considered to be in ``quasiequilibrium''.
When this condition is satisfied we are allowed to drop the $\Phi_{,tt}$ 
(radiative) term from the field equation. Doing this the problem reduces
to the solution of three ordinary differential equations which can be 
solved ``analytically''. 
We will call this simpler problem the ``static'' approximation to the 
exact problem, 
\bq \label{st-problem}
\left\{
\begin{array}{l}
\nabla^2 \Phi=\mbox{source}~~,\\
\ddot{r}_p=\ldots~~,\\
\ddot{\phi}_p=\ldots~~.
\end{array}
\right.
\eq
In the static approximation (which reduces to Newtonian gravity in the 
limit $\alpha\gg 1$) the particle motion is conservative but not 
necessarily periodic due to the nonlinearity of the problem.

\begin{figure}[hbt]
\begin{center}
\includegraphics[width=10cm]{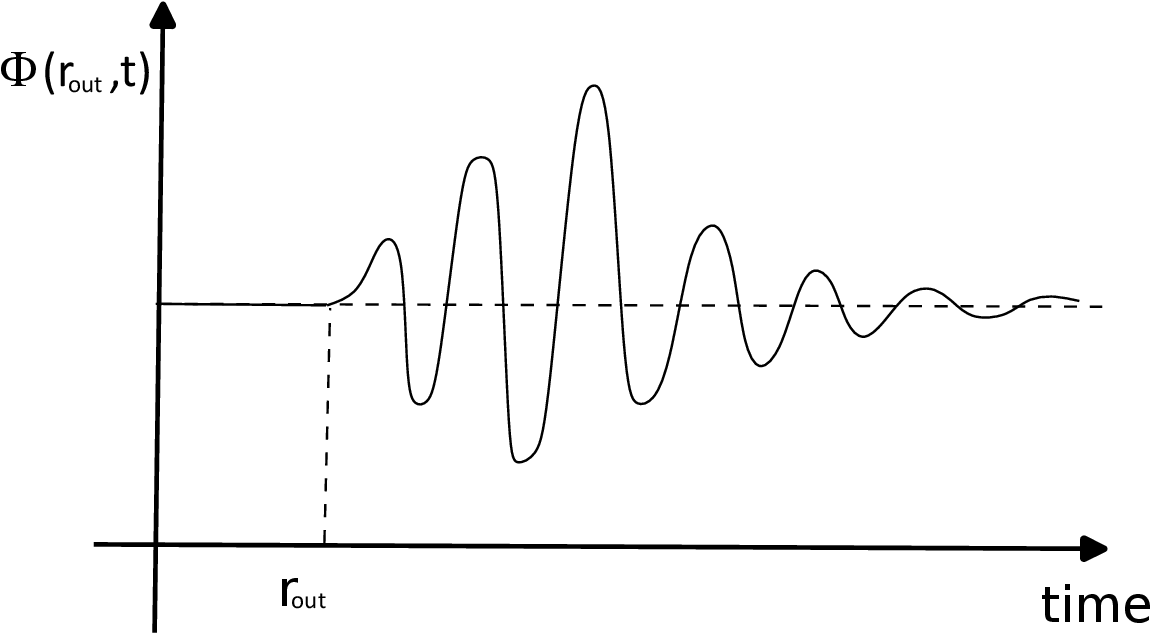}\\
\end{center}
\caption{Expected behaviour for $\Phi(r_{out},t)$ as a function of time.
\label{true-wave}}
\end{figure}

\begin{figure}[hbt]
\begin{center}
\includegraphics[width=10cm]{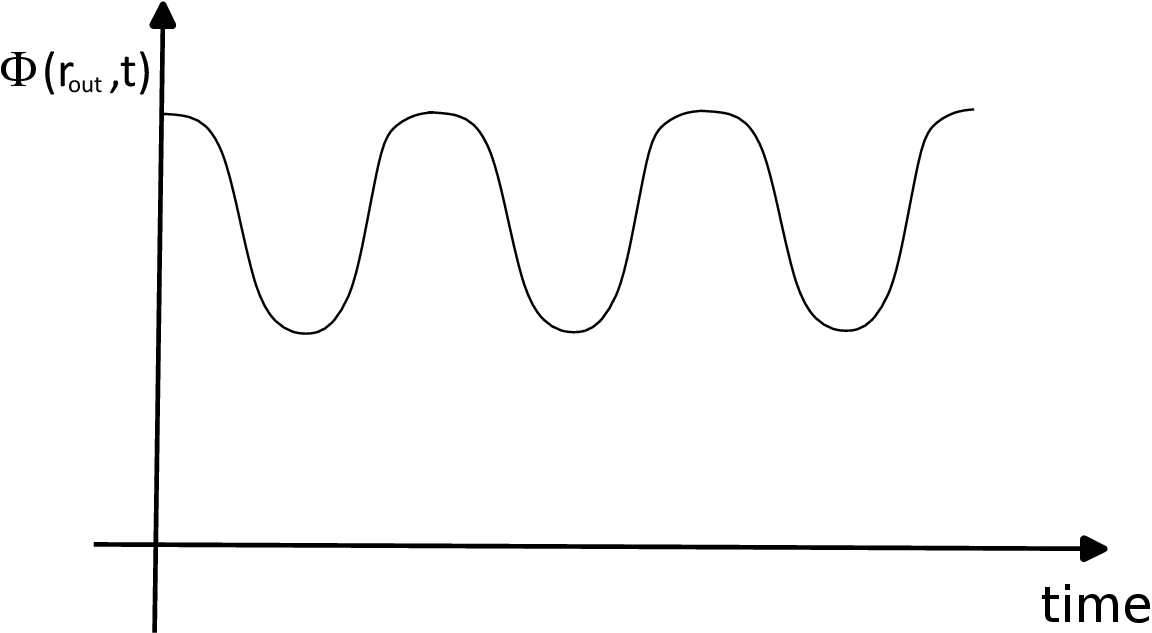}\\
\end{center}
\caption{Same as figure \ref{true-wave} but in the static approximation.
\label{static-wave}}
\end{figure}

Monitoring the exact solution for the field at a fixed radius $r_{out}$ 
far from the particle, one expects a behaviour similar to the one
shown in figure \ref{true-wave}.
In particular the damping of the wave amplitude is due to the fact that 
the particle is gradually approaching a circular orbit.
In the static approximation the field cannot have any damping because 
of the conservativeness of the particle motion, and we get 
a behaviour as shown in figure \ref{static-wave}.

Any reasonable approximation to the exact solution in the nonlinear regime
has to be able to reproduce the damping of the wave. The ``quasistatic'' 
approximation that we propose takes into account the wave damping 
through the following four steps:
\begin{description}
\item[(1)] We use the solution $r_p(t)$ to the static approximation
to determine the field equation source term. We then solve the full
field equation,
\bq
\square \Phi(\rr,t)=\mbox{source}~~,\\
\eq
to find the flux of field energy ($\sim r^2\Phi_{,t}\Phi_{,r}$) radiated
out by the gravity wave. This will allow us to determine the rate of 
change of the total energy $E$ of the particle-field system, with respect
to time,
\bq
\frac{dE}{dt}=-\int \mbox{flux} \; da~~.
\eq
\item[(2)] Consider the particle-field system in the stationary state where
the particle is in a circular orbit at a radius $R$. Then instantaneously
change the particle angular momentum from $J=1$ to 
$J=\utlp/(\utlp)_{circ}(R)$ and calculate the total energy of the system.
Repeating this for all radii $R$ between $r_i$ and $r_e$ we get a curve
$E(R)$ similar to the one shown in figure \ref{E(R)}. The values 
$E(r_e)$ and $E(r_i)$ are exact, while at the true inversion points 
$r_{inv}$ of the particle orbit, $E(r_{inv})$ are expected to be good 
approximations to the corresponding exact values. Knowing $E(R)$ we can
find the rate of change of $E$ with respect to $R$. 
\begin{figure}[hbt]
\begin{center}
\includegraphics[width=10cm]{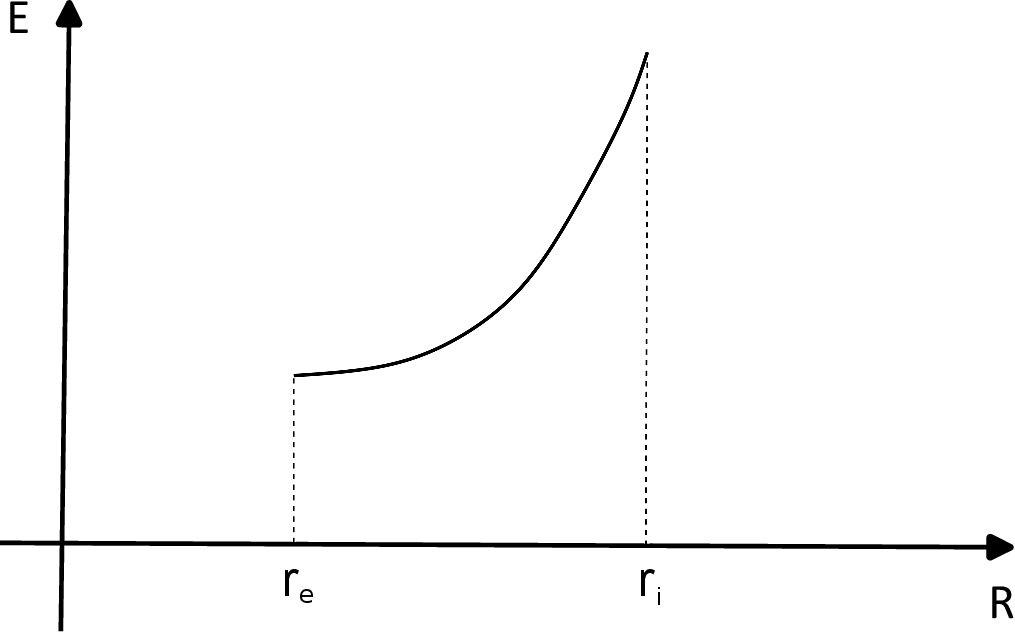}\\
\end{center}
\caption{Shows the expected behaviour for the total energy of the system
$E$ as a function of the circular orbits radii $R$. The energy curve has
its minimum at $r_e$, the radius of the circular orbit on which the 
particle decays at $t=\infty$.
\label{E(R)}}
\end{figure}
\item[(3)] Use the chain rule to get the rate of change of $R$ with 
respect to time,
\bq
\frac{dR}{dt}=\frac{dE/dt}{dE/dR}~~.
\eq
\item[(4)] Finally knowing $dR/dt$ we can correct the previous static 
estimate of the field equation source term. We can then solve the full 
field equation again to get the wave damping.
\end{description}

\chapter[Statement of the problem]{Statement of the problem}
\label{prob}

\section[Basic Equations]{Basic Equations}

The gravitational field is described by a massless scalar field
$\Phi(x^\alpha)$ in special relativity. The scalar field does 
not modify the background space-time geometry which is always 
Minkowskian. Consider a particle of rest mass $m$ moving along a 
world-line $z^\alpha(\lambda)$. Then the action for the 
field-particle system is,
\bq
I=\int \mbox{$\cal L$}\:(-g)^{1/2}\:d^4x~~,
\eq
where the lagrangian density $\cal L$ is,
\bq \label{lagrangian}
\mbox{$\cal L$}=-\frac{1}{8\pi G}\:g^{\alpha\beta}\Phi_{,\alpha}
\Phi_{,\beta}-\rho e^{\Phi}~~,
\eq
and where the comoving density is,
\bq
\rho=m\int\left(-g_{\alpha\beta}\frac{dz^\alpha}{d\lambda}
\frac{dz^\beta}{d\lambda}\right)^{1/2}\delta^4(\vec{x}-\vec{z}(\lambda))
\:(-g)^{-1/2}\:d\lambda~~.
\eq 
Here the metric tensor $g_{\alpha\beta}$ is the usual Minkowski
metric $\eta_{\alpha\beta}$ since space-time is flat in this theory
[i.e. $g_{\alpha\beta}=\eta_{\alpha\beta}=\mbox{diag}(-1,1,1,1)$ in 
Cartesian coordinates] and $g=\det\{g_{\mu\nu}\}=-1$. 
We use arrows to denote four-vectors and boldface to 
denote three-vectors. We will set the speed of light $c=1$ but will 
display the gravitational coupling constant (Newton' s constant) $G$
explicitly. If we choose $\lambda$ equal to the proper time $\tau$
along the particle world-line, then,
\bq
\rho&=&m\int\delta^4(\vec{x}-\vec{z}(\tau))\:(-g)^{-1/2}\:d\tau\\
    &=&\frac{m}{\gamma}\delta^3({\bf x}-{\bf z}(t))\:(-g)^{-1/2}~~,
\eq
where $\gamma\equiv dz^0/d\tau$ is the Lorentz factor.

Varying the Lagrangian (\ref{lagrangian}) with respect to $\Phi$ gives
the field equation of motion,
\bq \label{fe}
\square\Phi=4\pi G e^\Phi \rho~~.
\eq
In the Newtonian limit, where $\Phi \ll 1$, equation (\ref{fe}) 
becomes linear and reduces to Poisson' s equation. Varying the 
lagrangian with respect to $z^\alpha$ gives the particle equation 
of motion,
\bq \label{particle}
\frac{D^2z^\alpha}{d\tau^2}+\left[ g^{\alpha\beta}+
\frac{dz^\alpha}{d\tau} \frac{dz^\beta}{d\tau}\right] \Phi_{,\beta}
=0~~,
\eq 
where $D$ denotes covariant differentiation. Here we are allowing for 
curvilinear coordinates; covariant differentiation reduces to ordinary
differentiation in Cartesian coordinates. In the non relativistic
limit equation (\ref{particle}) implies that the gravitational force 
is $-\nabla\Phi$. The fully relativistic form ensures that the 
four-velocity $u^\alpha=dz^\alpha/d\tau$ remains orthogonal to the 
four-acceleration $a^\alpha=Du^\alpha/d\tau$ 
(in fact $u_\alpha u^\alpha=-1$).

\section[The problem]{The problem}

Because scalar gravitation can generate gravitational
waves in spherical symmetry, we can test out computational
algorithms for calculating gravitational radiation
on {\sl one-dimensional} systems. This is much simpler than
in general relativity theory. While not every aspect of the
general relativistic problem is reflected in this setting,
many features of wave generation in a rapidly varying
nonlinear dynamical system are exhibited here.

Consider one particle of rest mass $m$ moving along a world-line 
$z^\alpha(\tau)=(\rr_p,t)$ with four-velocity $u^\alpha$, under the 
influence of a massless scalar gravitational field $\Phi(\rr,t)$ in 
special relativity. 
In spherical symmetry the comoving matter density takes the form,
\bq
\rho(r,t)=\frac{m/\gamma}{4\pi \rp^2(t)}\delta(r-\rp(t))~~,
\eq
where $r=|\rr|$ and $\gamma=u^0$ is the Lorentz factor. The particle
effectively represents an entire spherical shell of radius $\rp$ 
and mass surface density $\sigma=m/(\gamma 4 \pi\rp^2)$. 

Assuming the particle confined in the $\theta=\pi/2$ plane, so that 
$u^\theta=0$ at all times, the equations of motion in spherical 
coordinates $\rr_p=(r_p,\theta_p,\phi_p)$, are (see Appendix \ref{app:0}),
\bq \label{pe}
\left\{
\begin{array}{l}
\dot{r}_p=\frac{\utlr}{\utt} \\ \\
\dot{\tilde{u}}_r=\frac{\utlp^2}{\utt\:\rp^3}-\frac{e^{2\Phi}\Phi_{,r}}{\utt}
\end{array}
\right.,~~
\left\{
\begin{array}{l}
r_p^2\dot{\phi}_p=\frac{\utlp}{\utt}\\ \\
\dot{\tilde{u}}_\phi=0
\end{array}
\right.~~,
\eq
where the dot stand for a time derivative and, 
\bq \label{penergy}
&&\utt=\sqrt{e^{2\Phi}+\utlr^2+\utlp^2/\rp^2}~~,\\
&&\tilde{u}^\alpha\equiv e^\Phi u^\alpha~~,
\eq
and we use the dot to denote total differentiation with respect to time,
and commas to indicate partial differentiation.

The particle moves conserving its orbital angular momentum $\utlp$.
For a static gravitational field the particle energy $\utt$ is also 
a constant.

Notice that from the field equation (\ref{fe}) follows that
$\phi_{,r}$ has, at all times
\footnote{We can safely assume that $\Phi_{,tt}$ remains finite at all
times at the shell surface.}
, a jump of $4\pi G e^{\Phi}\sigma$ 
at the shell surface $r=\rp(t)$. It is then necessary to specify 
how we calculate the gravitational force felt by the shell. 
We will use in equation (\ref{pe}),
\bq
\Phi_{,r}\equiv[\Phi_{,r}(\rp-)+\Phi_{,r}(\rp+)]/2~~.
\eq
In this way we prevent any small patch of surface on the shell from
interacting with itself.

\section[Initial condition]{Initial condition}
\label{initial-condition}

The field starts from a moment of time symmetry, so that at $t=0$,
\bq \nonumber
&&\Phi_{,t}=0~~,\\
&&\nabla^2 \Phi=4\pi G e^\Phi \sigma \delta(r-r_i)~~,
\eq
where $r_i=\rp(t=0)$ is the initial shell radius.
The field is subject to the boundary conditions,
\bq
&&\Phi_{,r}=0~~~~~~~~~~~~r=0~~,\\
&&(r\Phi)_{,r}=0~~~~~~~~r\rightarrow \infty~~.
\eq
Choosing,
\bq \label{field0}
\Phi=\left\{
\begin{array}{ll}
a/\rp & r\leq \rp\\
a/r   & r>\rp
\end{array}
\right.~~,
\eq
we can determine $a_i=a(t=0)$ from the matching condition at the shell' s
surface,
\bq \label{jump}
\Phi_{,r}(\rp+)-\Phi_{,r}(\rp-)=\frac{G m \:e^{2\Phi}}{\rp^2\utt}~~.
\eq
Initially the particle is in a circular orbit of radius $r_i$ around 
the origin,
\bq \nonumber
&& \utlr=0~~, \\
&& r_i (u^\phi_{circ})^2=[\Phi_{,r}(r_i-)+\Phi_{,r}(r_i+)]/2=
-\frac{a_i}{2r_i^2}~~, 
\eq
with an angular momentum,
\bq \label{utlpe}
(\utlp)_{circ}=e^\Phi r_i^2 u^\phi_{circ}=e^{\textstyle a_i/r_i}
\sqrt{-\frac{a_i r_i}{2}}~~.
\eq
We can then find $a_i$ from equation (\ref{jump}), which becomes,
\bq \label{a1}
a_i=-\frac{Gm\:e^{\textstyle a_i/r_i}}{\sqrt{1-a_i/(2r_i)}}~~.
\eq

This initial condition (an Einstein state) is a stationary wave 
for the field equation of
motion and a stable circular orbit for the particle. So if we let the
system evolve from this initial state nothing will happen: the
particle will keep moving in the circular orbit at
radius $\rp(t)=r_i$ under the influence of the static gravitational
field (\ref{field0}). This can be shown, for example, rewriting the 
field equation of motion in terms of the auxiliary functions,
\bq \nonumber
X(r,t)=[(r\Phi)_{,r}+(r\Phi)_{,t}]/2~~,\\
Y(r,t)=[(r\Phi)_{,r}-(r\Phi)_{,t}]/2~~.
\eq
Now equation (\ref{fe}) becomes,
\bq \nonumber
&& X_{,t}=X_{,r}-F\delta(r-\rp)~~,\\ \label{afe}
&& Y_{,t}=-Y_{,r}+F\delta(r-\rp)~~,
\eq
where $F=Gm\exp(2\Phi)/(2\rp\utt)$.
The initial condition for X and Y becomes,
\bq 
X(r,0)=Y(r,0)=\left\{
\begin{array}{ll}
a_i/(2r_i) & -r_i <r< r_i\\
0         & \mbox{otherwise} 
\end{array}
\right.~~,
\eq
From equations (\ref{utlpe}) and (\ref{a1}) follows that when $\utlp/(\utlp)_{circ}=1$, 
at $t=0$, the source term $F=a_i/(2r_i)$. So that after an infinitesimal 
timestep $dt$, $X(r,dt)=X(r,0)$ and $Y(r,dt)=Y(r,0)$.

So we perturb the system changing the particle' s angular momentum 
by a factor $\xi$,
\bq
\utlp=\xi\; (\utlp)_{circ} ~~,
\eq 
and let it evolve.

\section[Conserved integrals]{Conserved integrals}

The particle-field dynamical system is characterized by a time-varying matter
and velocity profile, interacting with a time varying scalar field
containing radiation. Conservation of energy-momentum follows from,
\bq \label{divT}
\nabla T=0~~,
\eq
where $T$ is the total stress-energy tensor of the system,
\bq
T^{\mu\nu}=\frac{2}{(-g)^{1/2}}\frac{\delta [\mbox{$\cal L$}(-g)^{1/2}
]}{\delta g_{\mu\nu}}~~.
\eq
Carrying out the variation using equation (\ref{lagrangian}) we find,
\bq
T_{\mu\nu}=T_{\mu\nu}^{field}+T_{\mu\nu}^{particle}~~,
\eq
where,
\bq
&&T_{\mu\nu}^{field}=\frac{1}{4\pi G}[\Phi_{,\mu}\Phi_{,\nu}-\frac{1}{2}
g_{\mu\nu}\Phi^{,\alpha}\Phi_{,\alpha}]~~,\\
&&T_{\mu\nu}^{particle}=\rho e^\Phi u_\mu u_\nu~~.
\eq

Conservation of energy-momentum  gives rise to the 
following conserved integrals,
\bq
\frac{\partial}{\partial t}\int_{S_r}T^{\mu 0}({\bf x},t)\:d^3x=
-\int {T^{\mu i}}_{,i}\:d^3x=-4\pi r^2T^{\mu r}(r,t)~~,
\eq
where $S_r$ is the volume of the sphere of radius $r$ centered at the
origin, and we used spherical symmetry in the last equality. 

\begin{figure}[hbt]
\begin{center}
\includegraphics[width=10cm]{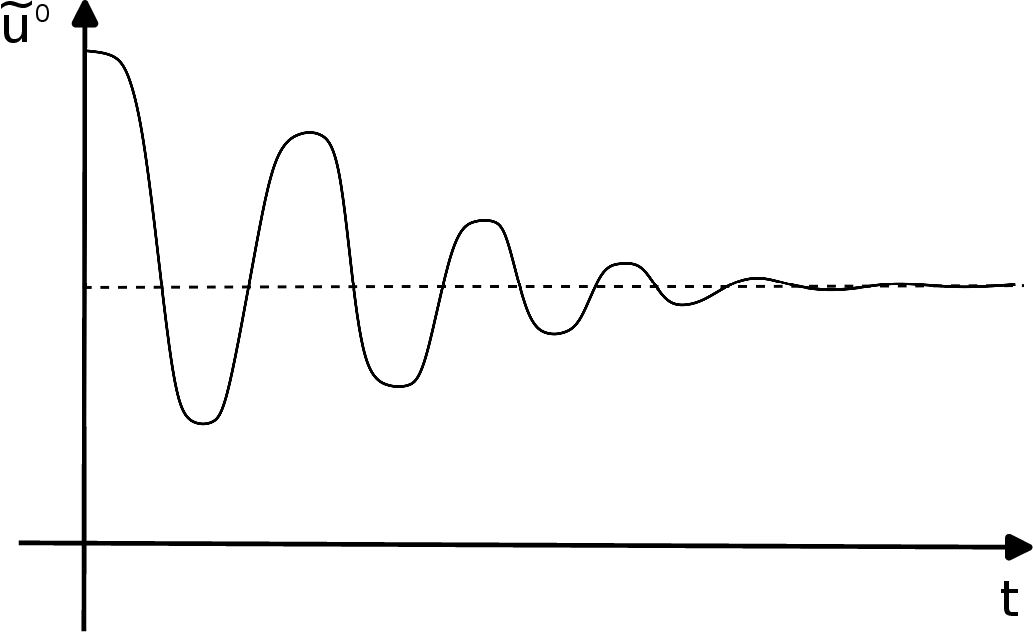}\\
\end{center}
\caption{Shows the expected behaviour of the particle energy $\utt$ as
a function of time for the case $\alpha\sim 1$ and $\xi<1$.
\label{utt}}
\end{figure}

When $r>r_p(t)$ we find,
\bq \label{m0a}
&[\mu=0]&\frac{\partial}{\partial t}\left\{ \frac{1}{2G}\int_0^r
[(\Phi_{,0})^2+(\Phi_{,r})^2]r^2\:dr+m\utt\right\}=\frac{1}{G}
r^2\Phi_{,0}\Phi_{,r}~~,\\ \nonumber
&[\mu=\phi]&\frac{\partial}{\partial t}\left(\frac{\utlp}{\rp^2}\right)=0
~~,\\ \nonumber
&[\mu=r]&\frac{\partial}{\partial t}\left\{ \frac{1}{G}\int_0^r
\Phi_{,0}\Phi_{,r} r^2\:dr-m\utlr\right\}=\frac{1}{2G}
r^2[(\Phi_{,0})^2+(\Phi_{,r})^2]~~,
\eq
The particle-field total mass energy is given by the integral in
equation (\ref{m0a}),
\bq \label{tote}
&&E=E^{field}+E^{particle}~~,\\ \nonumber 
&&E^{field}=\frac{1}{2G}\int_0^r [(\Phi_{,0})^2+(\Phi_{,r})^2]r^2\:dr
~~,\\ \nonumber
&&E^{particle}=m\utt~~.
\eq  
According to equation (\ref{m0a}), when evaluated at large enough
radius, outside any radiation or matter, $E$ is
conserved. As the particle shell breaths around its asymptotic virial
equilibrium state, $E^{particle}$ will undergo exponentially dumped 
oscillations around its asymptotic value (see figure \ref{utt}): 
the oscillations are due to the coupling
with the field, and the dumping to the gravitational radiation
going out to infinity (as a gravity wave). 
So that after a long time, apart from some
particular combinations of $\xi$ and $r_i/m$ (see section \ref{relax}), 
some energy will have been exchanged between the field and the particle.

For a static situation, the $(\Phi_{,r})^2$ term in equation (\ref{tote})
can be integrated by parts to get,
\bq \label{static-energy}
E=-\frac{m\Phi_p e^{2\Phi_p}}{2\utt}+m\utt~~,
\eq
where $\Phi_p=\Phi(r_p,t)$. In the Newtonian limit equation 
(\ref{static-energy}) becomes,
\bq \nonumber
E&=&m\left[-\frac{\Phi_p}{2}+\cdots+(1+\Phi_p+\cdots)\left(1+\frac{v_r^2}{2}+
\frac{v_{\phi}^2}{2r^2}+\cdots\right)\right]\\
&\approx& m\left(1+\frac{v_r^2}{2}+\frac{v_{\phi}^2}{2r^2}+
\frac{\Phi_p}{2}\right)~~,
\eq
where $v_i\equiv u_i/u^0$. So $E$ is the sum of the rest mass, plus the 
kinetic energy, plus the gravitational potential energy of the matter 
shell.

When $r<r_p(t)$,
\bq
&[\mu=0]&\frac{\partial}{\partial t}\int_0^r[(\Phi_{,0})^2+
(\Phi_{,r})^2]r^2\:dr=2r^2\Phi_{,0}\Phi_{,r}~~,\\
&[\mu=\phi]& 0=0~~,\\
&[\mu=r]&\frac{\partial}{\partial t}\int_0^r\Phi_{,0}\Phi_{,r} r^2
\:dr=\frac{r^2}{2}[(\Phi_{,0})^2+(\Phi_{,r})^2]~~,
\eq
which implies,
\bq
(\Phi_{,0})^2=(\Phi_{,r})^2~~~~\forall t~~~~\forall r<\rp(t)~~.
\eq

\begin{figure}[hbt] 
\begin{center}
\includegraphics[width=10cm]{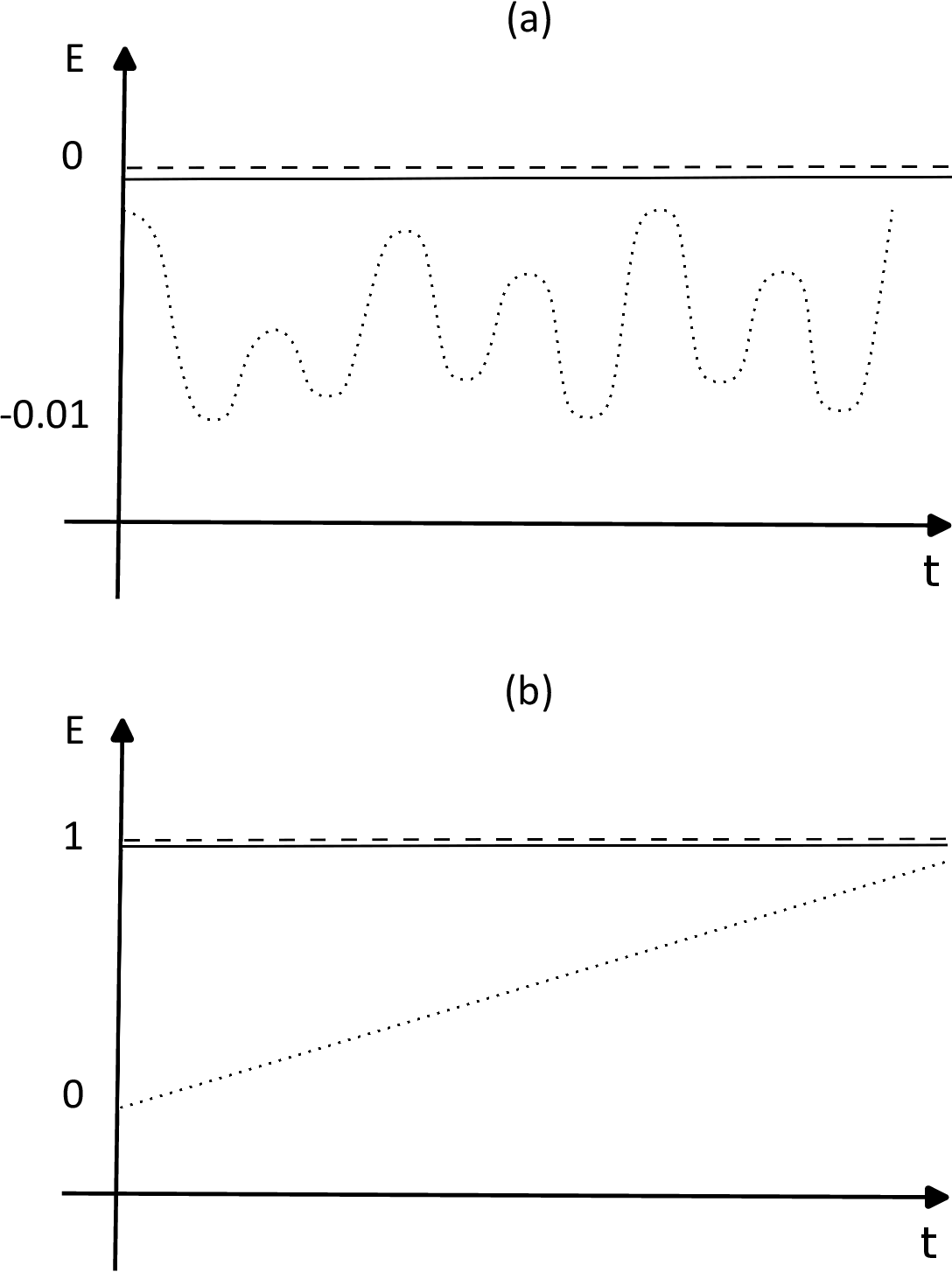}\\
\end{center}
\caption{Energy conservation at two selected radii as a function of
time. The solid line shows the left-hand side of equation (\ref{ec})
(volume integral plus integrated flux), the dotted line shows the
second term alone (integrated flux), and the dashed line shows the
right-hand side (volume integral at $t=0$). The radii are (a)
$r_{ec}<r_p(t)$ at all times, (b) $r_{ec}>r_p(t)$ at all times. 
The degree to which the solid and dashed lines coincide compared with
the magnitude of the dotted line is a measure of the code' s ability
to conserve energy.
\label{econs}}
\end{figure}

Those conserved integrals can be used as self consistent checks on our
numerical integration. In figure \ref{econs} we show what we would
expect if we were to evaluate the energy conservation equation,
\bq \nonumber
\int_0^{r_{ec}}[(\Phi_{,0})^2+(\Phi_{,r})^2]r^2\:dr+2m\utt
\theta(r_{ec}-r_p(t))\\ \nonumber
-2\int_0^t dt[\Phi_{,0}\Phi_{,r}r_{ec}^2-m\delta(r_{ec}-r_p(t))\utlr]=
\\ \label{ec}
\int_0^{r_{ec}} (\Phi(r,0)_{,r}r)^2 \:dr+2m\utt(t=0)\theta(r_{ec}-r_p(0))~~.
\eq
as a function of time at two fixed radii $r_{ec}$. The first radius is
inside the shell at all times, the second is always in the vacuum
exterior. In the first case the right hand side of equation (\ref{ec})
is zero, the integrated flux term (second integral in equation
(\ref{ec})) is large, and the energy conservation involves
the cancellation of large terms. Consequently, the high degree to
which we are able to maintain energy conservation is a nontrivial
measure of the accuracy of the code. In the exterior, the flux is
small and energy conservation is not a stringent test. 
\vspace{3cm}

\section[Monopole radiation]{Monopole radiation}
\label{monopole}

In the weak field, slow motion limit, the radiation field can be
obtained by a multipole expansion.
Since the theory involves a scalar field, the lowest-order
contribution to the radiation comes from the monopole term. This is in
contrast with electromagnetism (vector field: dipole radiation) or
general relativity (tensor field: quadrupole radiation).

Using Green' s function for the wave equation we can transform
equation (\ref{fe}) into the integral form,
\bq \label{integral}
\Phi(\xb,t)=-G\int d^3\xp \frac{[e^\Phi\rho]_{\tp=t-|\xb-\xbp|}}
{|\xb-\xbp|}~~. 
\eq
In the wave zone we can replace the denominator in equation
(\ref{integral}) by the distance $r=|\xb|$. To isolate the conserved
rest mass $m$, define the rest density to be,
\bq
\rho_0=\gamma\rho~~.
\eq 
Then,
\bq \label{rad1}
\Phi(\xb,t)\approx -\frac{G}{r}\int d^3\xp \left[\frac{e^\Phi}{\gamma}
\rho_0\right]_{\tp=t-|\xb-\xbp|}~~.
\eq 
In the integrand, expand,
\bq \label{rad2}
\rho_0(\xbp,\tp)=\rho_0(\xbp,t-r)+(r-|\xb-\xbp|){\rho_0}_{,t}\\ \nonumber
+\shalf(r-|\xb-\xbp|)^2{\rho_0}_{,tt}+\cdots~~,
\eq 
and,
\bq \label{rad3}
\frac{e^{\Phi}}{\gamma}=[1+\Phi-\shalf v^2]_{\tp=t-r}+\cdots~~,
\eq
where $v^2=[u_r/u^0]^2+[u_\phi/(u^0r)]^2$. For large $r$,
\bq
r-|\xb-\xbp|\approx\frac{\xb\cdot\xbp}{r}=\rpp \cos\tep~~.
\eq
The leading-order contribution to the expansion of equation
(\ref{rad1}) comes from the product of $\rho_0$ in equation
(\ref{rad2}) with the 1 in equation (\ref{rad3}). The resulting
integral gives $m$, so that this term represents the nonradiative
Coulomb field. Thus the leading-order radiation field is,
\bq \label{rad4}
\Phi(\xb,t)=-\frac{G}{r}\int d^3\xp [\rho_0(\Phi-\shalf v^2)+
\rpp\cos\tep{\rho_0}_{,t}+\rpp^2\cos^2\tep{\rho_0}_{,tt}]~_{t-r}
~~.
\eq
To this order, it is irrelevant wether one uses $\rho$ or $\rho_0$ in
equation (\ref{rad4}).

\begin{figure}[hbt] 
\begin{center}
\includegraphics[width=10cm]{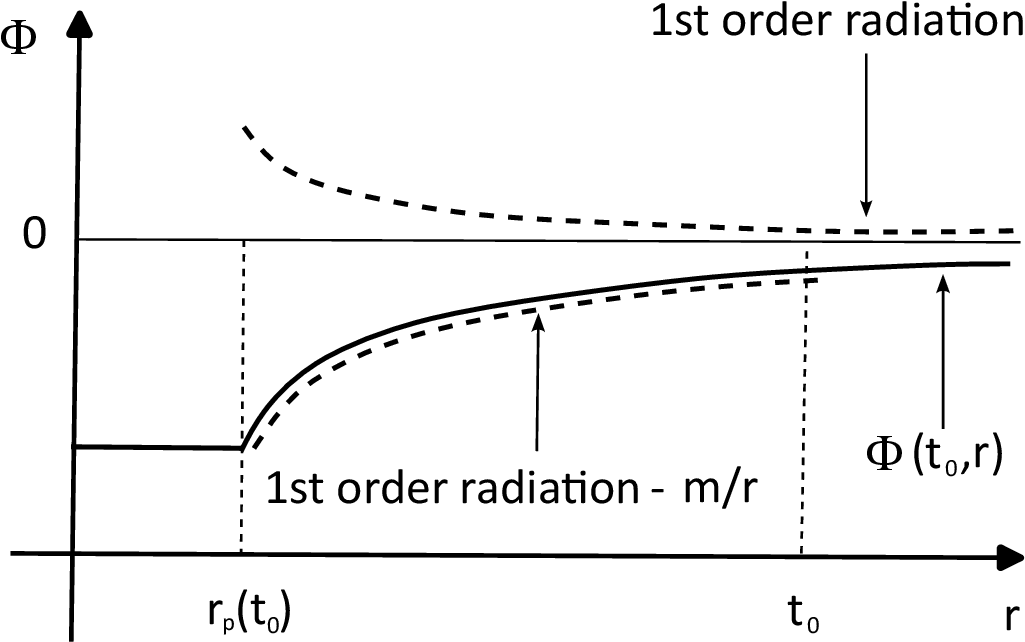}\\
\end{center}
\caption{For the case $\alpha\sim 1$, $\xi <1$, shows a snapshot at
$t=t_o$ of the field $\Phi(t_o,r)$, the first order radiation part
(\ref{wfsm}), and the first order radiation part plus the zeroth order
$-m/r$. 
\label{wave}}
\end{figure}

For a spherically symmetric density distribution, the term
proportional to $\cos\tep$ in equation (\ref{rad4}) integrates to
zero, giving,
\bq \label{srad}
\Phi(r,t)=-\frac{G}{r}\int d\rpp\: 4\pi\rpp^2[\rho_0(\Phi-\shalf v^2)
+\ssixth \rp^2{\rho_0}_{,tt}]_{t-r}~~.
\eq
The last term in the integrand can be rewritten as follows,
\bq
\frac{1}{6}\int d\rpp\: 4\pi\rpp^2\frac{d^2(\rpp^2)}{dt^2}\rho_0=
\frac{1}{3}m\left(\utlr^2+\rp\frac{d\utlr}{dt}\right)~~,
\eq
and using the equation of motion (\ref{pe}) in the weak field 
limit, we find,
\bq
\frac{1}{6}\int d\rpp\: 4\pi\rpp^2\frac{d^2(\rpp^2)}{dt^2}\rho_0=
\frac{1}{3}m\left(\utlr^2+\frac{\utlp^2}{\rp^2}-\rp\Phi_{,r}\right)~~,
\eq
Thus equation (\ref{srad}) becomes,
\bq \label{wfsm}
r\Phi(r,t)=Gm\left\{\frac{1}{6}\left[\rp(\Phi_{,r}(\rp+)+
\Phi_{,r}(\rp-))+\utlr^2+\frac{\utlp^2}{\rp^2}\right]-\Phi\right\}
_{t-r}~~.
\eq
In figure \ref{wave} we show how a snapshot of the field at $t=t_o$
should look like, and compare it with the leading order radiation
field of equation (\ref{wfsm}), in the wave zone.

From equation (\ref{m0a}) follows that the rate of energy emission
when $r>r_p(t)$ is,
\bq \label{erate}
\frac{dE}{dt}=-\frac{1}{G}r^2\Phi_{,t}\Phi_{,r}=
-\frac{1}{G}(r\Phi_{,t})^2~~,
\eq
where in the last equality we used the fact that since $X$ is
propagating to the left, the following outgoing wave boundary
condition must hold,
\bq
X(r,t)=0~~~~\mbox{or}~~~~(r\Phi)_{,r}+(r\Phi)_{,t}=0~~~~\forall t,
~~\forall r>\rp(t)~~.
\eq

\section[Analytic results]{Analytic results}

\subsection[Newtonian limit]{Newtonian limit}

\label{at}

For weak fields and slow velocities we can test our code using the
analytic solution from Newtonian gravitation. In this limit the
particle equation of motion is,
\bq
&&\ddot{\rp}=-\Phi_{,r}+\frac{J^2}{\rp^3}~~,\\ \nonumber
&&\Phi_{,r}=\frac{Gm}{2\:r^2}~~,\\ \nonumber
&&J=(r^2v_\phi)_{t=0}=r_i^2\xi\sqrt{\Phi_{,r}(r_i)/r_i}=\xi
\sqrt{Gmr_i/2}~~,
\eq
which can be rewritten as,
\bq
&&\ddot{x}=-\frac{M_{eff}}{x^2}+\frac{J_{eff}^2}{x^3}~~,\\ \nonumber
&&\dot{x}(0)=0~~,\\ \nonumber
&&x(0)=1~~, 
\eq
where $\rp(t)=r_ix(t)$, $M_{eff}=Gm/(2r_i^3)$, and $J_{eff}=\xi
\sqrt{M_{eff}}$. The first integral yields the conserved energy,
\bq
E=\frac{1}{2}\dot{x}^2-\frac{M_{eff}}{x}+\frac{J_{eff}^2}{2x^2}~~.
\eq
For $E=M_{eff}(\xi^2/2-1)<0$ (i.e. $\xi^2<2$) we have bound
orbits. Solving for the turning points ($\dot{x}=0$) yields,
\bq
x_{\pm}=\frac{1\pm(1-\xi^2)}{2-\xi^2}~~.
\eq
So that for $0<\xi<1$ the shell contracts to $r_i x_-$ and for
$1<\xi<\sqrt2$ it expands to $r_i x_-$. For $\xi>\sqrt 2$ the shell
explodes.

Integrating the equation of motion we get the parametric solution,
\bq 
&&x=a(1-e\cos(u))~~,\\ \nonumber
&&t=\frac{P}{2\pi}(u-e\sin(u))-\frac{P}{2}~~,
\eq
where the semimajor axis, eccentricity and period are given by,
\bq \nonumber
&& a=\frac{1}{2-\xi^2}~~,\\ \nonumber
&& e=|1-\xi^2|~~,\\ \label{period}
&& P=2\pi\sqrt{\frac{2r_i^3}{Gm(2-\xi^2)^3}}~~.
\eq

Inserting this analytic solution into equation (\ref{srad}) and
differentiating with respect to time gives the wave amplitude in the
wave zone,
\bq \label{eloss}
r\Phi_{,t}=-\frac{4}{3}\frac{(Gm)^2}{r_i}\left[\frac{\dot{x}}{x^2}
\right]_{t-r}~~.
\eq 
From equation (\ref{erate}) we get for the rate of energy emission,
\bq \label{Nerate}
\frac{dE}{dt}=-\frac{16}{9}\frac{(Gm)^4}{r_i^2}\left[\frac{\dot{x}^2}
{x^4}\right]_{t-r}~~.
\eq 
Integrating over an oscillation period we get the energy radiated per
period,
\bq \label{deltaE}
\Delta_PE=-\frac{16\pi}{36}m\left(\frac{Gm}{r_i}\right)^{7/2}
\frac{(1-\xi^2)^2}{\xi^7}(5-2\xi^2+\xi^4)~~.
\eq

\subsection[Relaxation to virial equilibrium]{Relaxation to virial
equilibrium}
\label{relax}

\begin{figure}[hbt] 
\begin{center}
\includegraphics[width=10cm]{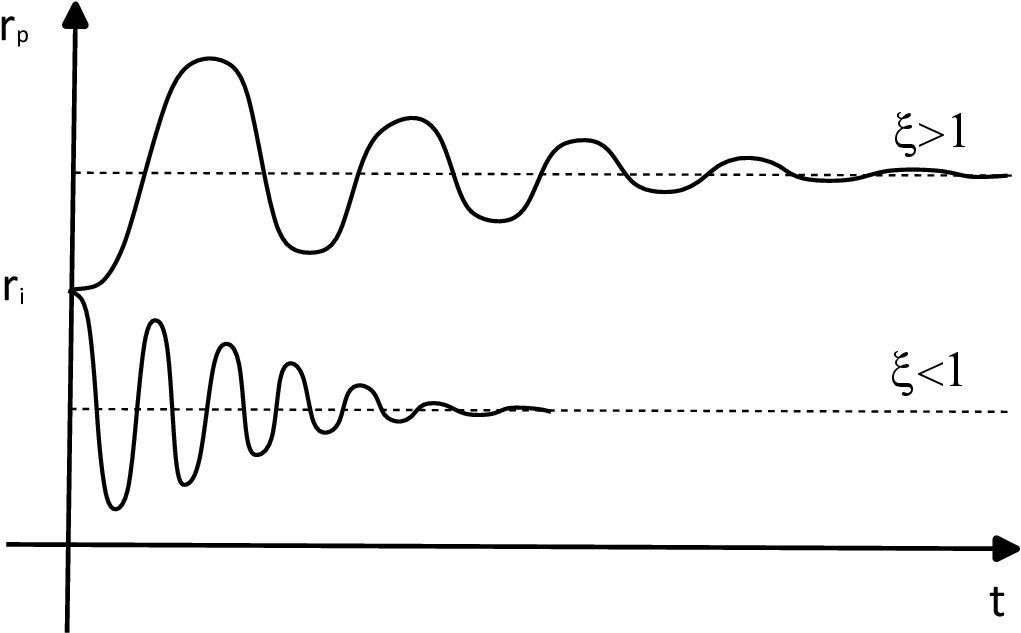}\\
\end{center}
\caption{Shows the relaxation to the virial equilibrium state for an 
$\alpha\sim 1$ shell with two different values of $\xi$: $\xi<1$ and
$\xi>1$. 
\label{behave}}
\end{figure}

If the shell does not explode or collapse,
it will eventually reach, as it loses energy by emitting gravitational
waves, an equilibrium circular orbit (see figure \ref{behave}). 
At this point the particle-field system is
in an Einstein state were $\utlr=0$, $\utlp^2=\rp^3
e^{2\Phi}\Phi_{,r}$, and the field is static and of the form (\ref{field0}),
in a neighborhood of the shell.

Given the angular momentum of the particle $\utlp$ we can then
predict the final equilibrium radius $r_e$, by solving the following
equations in $a_e=a(t=\infty)$ and $r_e=r_p(t=\infty)$,
\bq
&&a_e=-\frac{Gm\:e^{2a_e/r_e}}{\sqrt{e^{2a_e/r_e}+\utlp^2/{r_e}^2}}~~,\\
\label{eq}
&&\utlp^2=e^{2a_e/r_e}{r_e}^3(-a_e/(2{r_e}^2))~~.
\eq   
One can verify that,
\bq
\left\{
\begin{array}{ll}
r_e<r_i & \mbox{when}~~~\xi<1\\
r_e>r_i & \mbox{when}~~~\xi>1 
\end{array}
\right.~~,
\eq
 
This final state is a virial equilibrium state. Taking the trace of the 
special relativistic virial theorem,
\bq
\int T^{ij}d^3x=\frac{1}{2}\frac{\partial^2}{\partial t^2} \int T^{00}
x^i x^j d^3 x~~,
\eq
gives at equilibrium,
\bq
\int \rho e^\Phi ({u_r}^2+{u_\phi}^2/r^2)d^3 x=\frac{1}{8\pi G}\int(\nabla
\Phi)^2d^3x=-\frac{1}{2}\int\rho\Phi e^\Phi d^3x~~,
\eq 
or, when $\utlr=0$,
\bq
\frac{\utlp^2}{r_e^2}=-\frac{1}{2}e^{2\Phi}\Phi~~,
\eq
which is the same as equation (\ref{eq}), when the field is of the form
(\ref{field0}). 

The final energy of the particle-field system is,
\bq \label{einf}
E(t=\infty)=-\frac{m}{2}\frac{a}{r_e}\frac{e^{\textstyle 2a/r_e}}
{\utt(t=\infty)}+m\utt(t=\infty)~~,
\eq
where, 
\bq
\utt(t=\infty)=\sqrt{e^{\textstyle 2a_e/r_e}+\frac{\utlp^2}{r_e^2}}~~. 
\eq

The shell will only collapse into the origin when it possesses 0 angular 
momentum
\footnote{This is different from what happens in General Relativity 
where the shell can collapse also for non-zero values of the angular
momentum.}. 
This follows from equation (\ref{penergy}) and the observation that the 
particle energy $m\utt$ will always be smaller than the initial total 
energy of the particle-field system $E(t=0)$.
Since the exponential is bigger than 0, we can write,
\bq \label{lb}
[E(t=0)]^2>\utlr^2+\frac{\utlp^2}{r^2}~~,
\eq 
and when $\utlr=0$, equation (\ref{lb}) gives the following lower bound 
on the accessible radii
\footnote{At sufficiently small $\xi$ one can get a better lower bound
by substituting $E(t=0)$ with $\utt(t=0)$ in equation (\ref{rlb}).  
}, 
\bq \label{rlb}
\rp >\utlp/E(t=0)~~,
\eq 
where,
\bq \label{e0}
&&E(t=0)=\frac{a^2}{2r_i}+m\utt(t=0)~~,\\ \nonumber
&&\utt(t=0)=\sqrt{e^{\textstyle 2a/r_i}+\frac{\utlp^2}{r_i^2}}~~. 
\eq

\subsection[Explosion]{Explosion}

\begin{figure}[hbt] 
\begin{center}
\includegraphics[width=10cm]{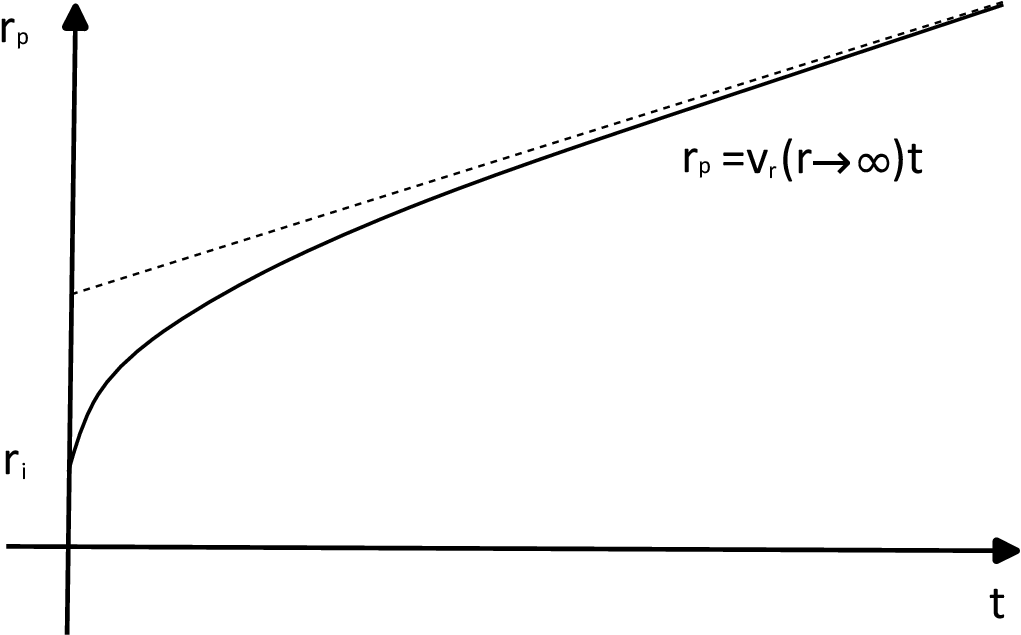}\\
\end{center}
\caption{Expected shell behaviour for $\xi>\xi_e$.
\label{expl}}
\end{figure}

In order to explode the shell has to reach $r=\infty$ with at least 
$\utlr=0$. But for $r\rightarrow \infty$, $\sigma\rightarrow 0$  
and $\phi\rightarrow 0$ so that $E^{particle}\rightarrow m$. When 
the shell is at infinity $E^{field}$ will be a small positive 
quantity.
So for the explosion to happen the initial energy of the 
particle-field system has to be greater then $m$,
\bq \label{explosion}
E(t=0)>m~~,\\ \nonumber
\eq
In the Newtonian limit $r_i\gg m$, condition (\ref{explosion}) reduces 
to $\xi>\sqrt 2$. The escape radial velocity is (see figure \ref{expl}),
\bq \nonumber
\utlr(r\rightarrow\infty)&=&\sqrt{[(E(t=0)-E^{field}(t=\infty))/m]^2-1}\\
&\approx&\sqrt{[E(t=0)/m]^2-1} ~~,
\eq
or,
\bq
v_r(r\rightarrow\infty)\approx\frac{\sqrt{E^2(t=0)-m^2}}{E(t=0)}~~.
\eq

\chapter[Approximations]{Approximations}

Here we will describe two approximated solutions of the exact problem
stated in chapter \ref{prob}.

\section[Quasistatic approximation]{Quasistatic approximation}

\label{qsa}

When it takes many oscillations for the particle to settle into the final
stable circular orbit, we can hope to approximate its slow motion with
a quasistatic approximation. The idea is the following. Consider the
static version of our problem (equations (\ref{pe})-(\ref{fe})),
\bq \label{qs}
&&(r\Phi)_{,rr}=\frac{Gme^{2\Phi}}{\utt\:r_s}\delta(r-r_s)~~,\\ \nonumber
&&\frac{dr_s}{dt}=\frac{\utlr}{\utt}~~,\\ \nonumber
&&\frac{d\utlr}{dt}=\frac{\utlp^2}{\utt\:r_s^3}-\frac{e^{2\Phi}\Phi_{,r}}
{\utt}~~,\\ \nonumber
&&\utt=\sqrt{e^{2\Phi}+\utlr^2+\utlp^2/r_s^2}~~,\\ \nonumber
&&\utlp=\mbox{constant}~~, 
\eq
where we called $r_s(t)$ the shell radius in this static approximation. 
At all times the field must be of the form (\ref{field0}) with $a=a_s$.
Once we know $r_s(t)$ and $\utlr(t)$ we can determine the field 
from the jump condition (\ref{jump}),
\bq \label{nf}
a_s=-\frac{Gm\:e^{\textstyle 2a_s/r_s}}{\sqrt{e^{\textstyle 2a_s/r_s}+\utlr^2
+\utlp^2/r_s^2}}~~.
\eq
Since we have a static field $\Phi=\Phi(r,r_s(t),\utlr(t))$,
$\Phi_{,t}=0$ and the system is conservative. The shell will then
experience undumped oscillations around its final equilibrium radius
$r_e$.
In order to have the shell reach $r_e$, we need a recipe to dump the
oscillations. This will give us a quasistatic approximation to the
true shell motion. 

\begin{figure}[hbt]
\begin{center}
\includegraphics[width=10cm]{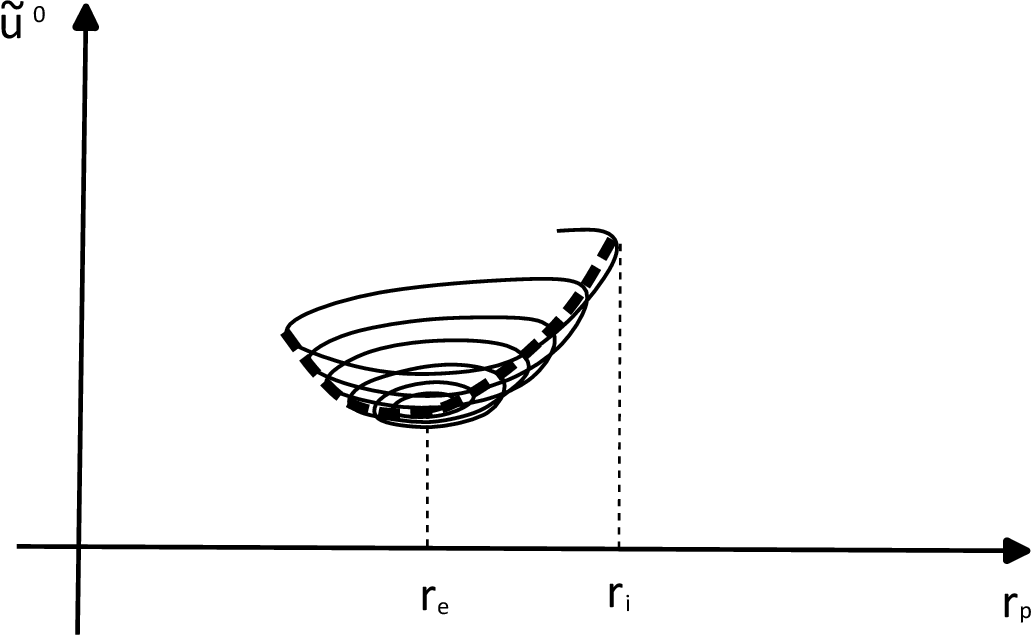}\\
\end{center}
\caption{For the $\alpha\sim 1$, $\xi<1$ case, shows $\utt$ as a function 
of the shell radius as expected from an exact numerical integration
(solid line) and from the analytic expression (\ref{uttqs}) (dashed line). 
We expect the dashed curve to pass through the true values for the
energy at the turning points of the particle orbit. 
\label{phsp2}}
\end{figure}

\begin{figure}[hbt]
\begin{center}
\includegraphics[width=10cm]{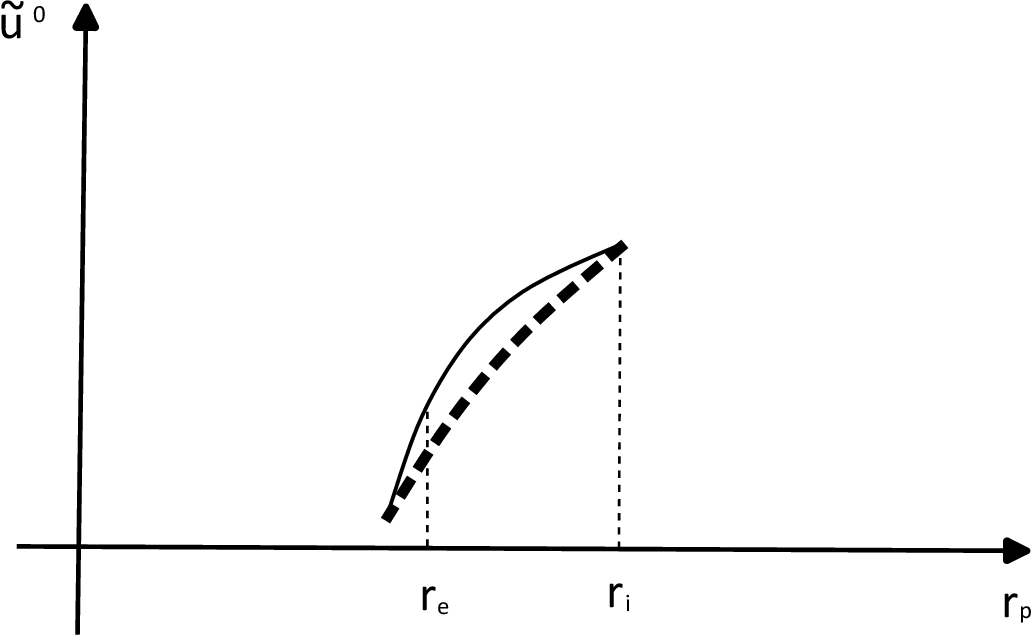}\\
\end{center}
\caption{Same as figure \ref{phsp2} but for the $\alpha \gg 1$ and
$\xi<1$ case. 
\label{phsp4}}
\end{figure}

\begin{figure}[hbt]
\centerline{\includegraphics[width=7cm]{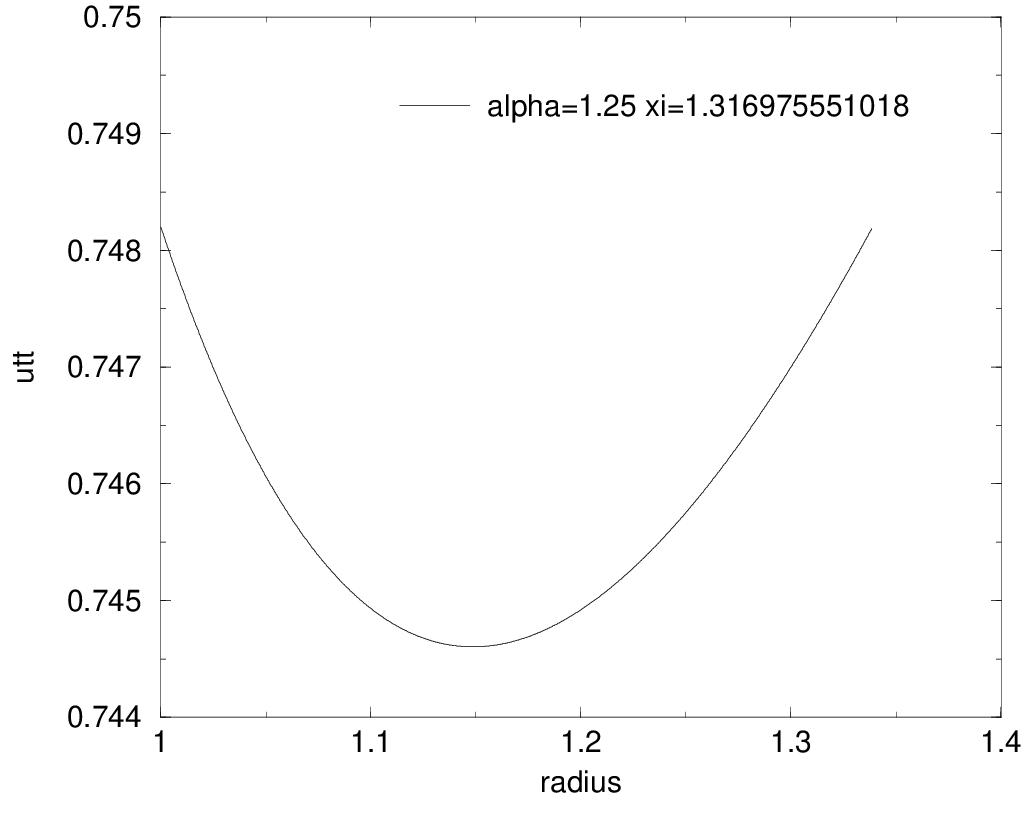}
\includegraphics[width=7cm]{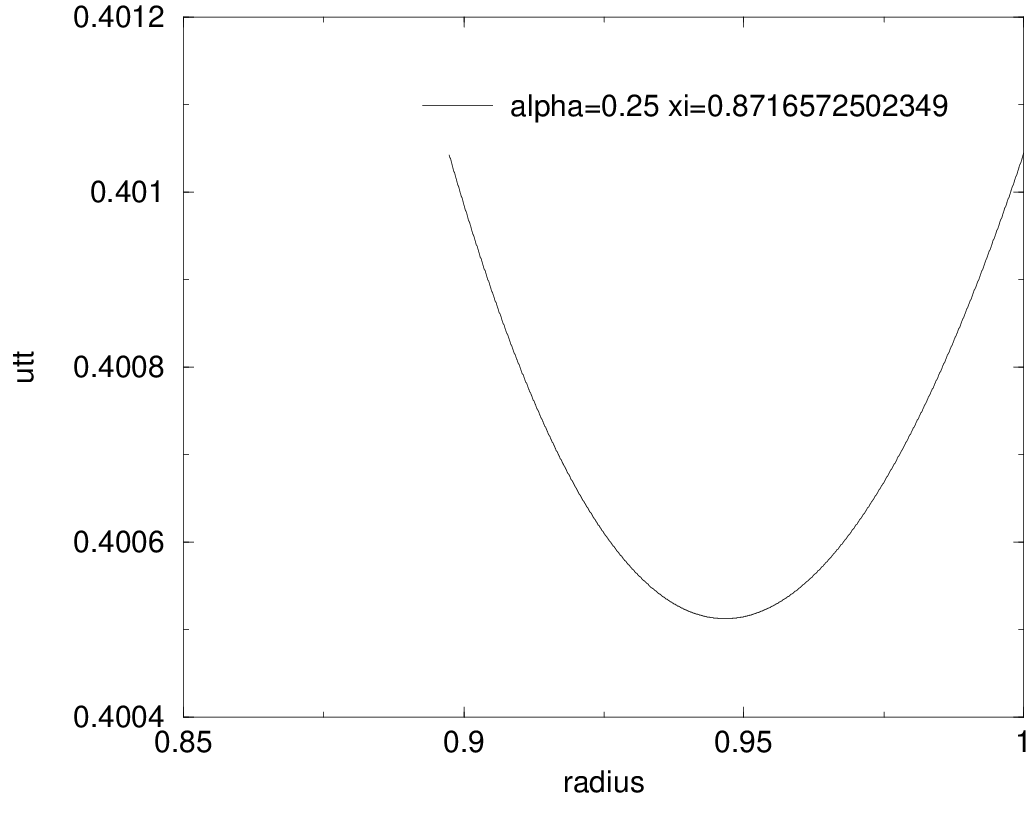}}
\caption{Shows $\utt$ as calculated from equation (\ref{uttqs}), as a 
function of the shell radius, for two different situations: on the 
left a more compact shell, on the right a less compact one. 
\label{utt12}}
\end{figure}

\begin{figure}[hbt]
\begin{center}
\includegraphics[width=10cm]{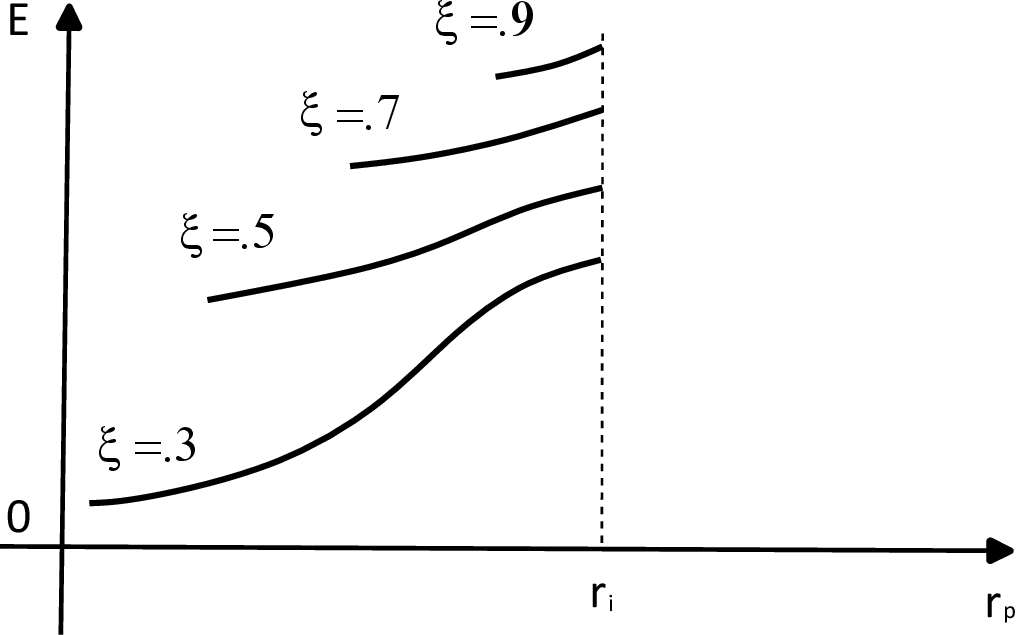}\\
\end{center}
\caption{Shows the expected family of curves for $E$ v.s. $r_p$
parametrized by the particle' s angular momentum $\xi$. 
\label{E-r}}
\end{figure}

Once we know the initial $r_i$, and final $r_e$ shell radii we can
construct a sequence of intermediate ``quasi-static'' Einstein states as
follows. The shell initially at $r_i$ will contract ($\xi<1$) or
expand ($\xi>1$) towards $r_e$ in a succession of circular orbits
($\utlr=0$) occurring at the true inversion points of the particle
trajectory. We call the intermediate radii of these circular orbits,
$\rqs(i)=\rp(P_1+\ldots+P_i)$, where $P_1,\ldots,P_i$ are the first
$i$ oscillation periods . At those points the field will be of the
form (\ref{field0}) with $a=\aqs$ determined by solving equation
(\ref{jump}) for any given $\rp=\rqs(i)$. At each $\rqs$ we can
determine the new value of $\xi$,
\bq
\xqs(i)=\frac{\utlp}{e^{\aqs/\rqs(i)}\rqs^2(i)\sqrt{-\aqs/(2\rqs^3(i))}}~~,
\eq
the particle energy,
\bq \label{uttqs}
\utt=\sqrt{e^{2\aqs/\rqs}+\utlp^2/\rqs^2}~~,
\eq 
and the particle-field energy,
\bq \label{qsE}
E=\frac{\aqs^2}{2\rqs}+m\utt
\eq
We expect this to be a very good approximation for the particle energy
at the true inversion points of the particle trajectory for a
wide range of $\alpha$'s and $\xi$'s (see figures \ref{phsp2} and
\ref{phsp4}). 
There usually is a value of $\xi$ different from 1, $\xi_o$, at which the 
energy of the particle in the final equilibrium state is equal to its 
energy at the beginning of the evolution (see figure \ref{utt12}). 
For shells with $\alpha<0.4204623... $, $\xi_o<1$, for less compact shells 
$\xi_o>1$. One can also show that $E(\rqs)$ has a minimum at
$\rqs(\infty)=r_e$ (see figure \ref{E-r}).

\begin{figure}[hbt]
\begin{center}
\includegraphics[width=10cm]{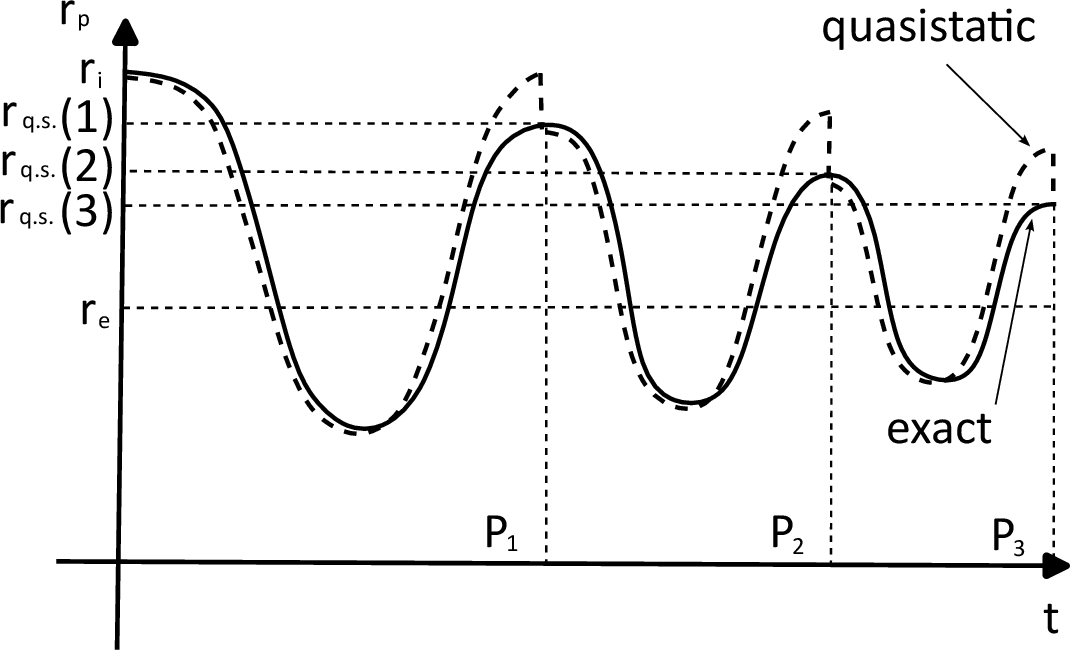}\\
\end{center}
\caption{How the quasistatic approximation is expected to approximate
a nonlinear collapse.
\label{qsapp}}
\end{figure}

Suppose we have approximated the true shell motion up to the i-th
period $P_i$. Then we continue the approximation as follows (see
figure \ref{qsapp}),
\begin{description}
\item[1.] Calculate $dE/d\rqs$ from equation (\ref{qsE}),
\bq \nonumber
\frac{dE}{d\rqs}&=&-\frac{\aqs^2}{2\rqs^2}+
\frac{\aqs^3(5\rqs-2\aqs)}{\rqs(7\aqs\rqs^2-2\aqs^2\rqs-4\rqs^3)}+\\
&&m\frac{\utlp^2}{\utt}\left(\frac{1}{\xqs^2}\frac{4\aqs-8\rqs}
{\rqs(7\aqs\rqs^2-2\aqs^2\rqs-4\rqs^3)}-\frac{1}{\rqs^3}\right)
\eq 
\item[2.] Calculate the energy radiated in the i-th oscillation period
$P_i$. In general, when $r>\rp(t)$, 
\bq \label{deltaE1}
\Delta_{P_i}E&=&\int_{P_1+\ldots+P_{i-1}}^{P_1+\ldots+P_i}dt\; 
\frac{dE}{dt}~~,\\
\frac{dE}{dt}&=&\frac{1}{G}r^2\Phi_{,t}\Phi_{,r} =
-\frac{1}{G}(r\Phi_{,t})^2~~.
\eq
We need a good approximation to the monopole term (the lowest order
contribution to the radiation) of the wave amplitude $r\Phi(r,t)$. In
the weak field slow motion limit one finds (see equation (\ref{srad})),
\bq \nonumber
r\Phi(r,t)=-G\int d\rpp\: 4\pi\rpp^2[\rho_0(\Phi-\shalf v^2)
+\ssixth \rp^2{\rho_0}_{,tt}]_{t-r}~~.
\eq
When $\alpha \gg 1$ it will be sufficient to use the Newtonian
approximation. So we will use the analytic expression (\ref{deltaE}). 
When $\alpha \sim 1$ we need to use the static approximation (system
(\ref{qs})) to get a numerical estimate for $\Delta_{P_i}E$. The
details of the calculation are outlined in the appendix. 

\item[3.] Given $\rqs(i)$ we can find $\rqs(i+1)$ using the chain rule,
\bq
\rqs(i+1)=\rqs(i)+\frac{\Delta_{P_i}E}{dE/d\rqs(\rqs(i))}~~.
\eq
\item[4.] Start a new static oscillation from $r_i=\rqs(i+1)$ and
$\xi=\xqs(i+1)$.  
\end{description}

\section[Characteristics approximation]{Characteristics approximation}
\label{cha}

We adopt a mean-field particle simulation scheme
\footnote{This scheme resembles Godunov's method used for the 
numerical solution of nonlinear systems of hyperbolic conservation 
laws \cite{Godunov59}
}:
\begin{description}
\item[1.] The particle is evolved in the mean background field $\Phi$
for a small time $\Delta t$. 
\item[2.] From the new particle position and velocity we obtain the
new matter source term appearing in the field equation (\ref{fe}). 
\item[3.] We then update $\Phi$ by evolving the field equation for a
time-step $\Delta t$.
\item[4.] Repeat the whole process.
\end{description}  
The particle evolves through an ordinary differential equation which
poses no computational difficulties. One can for example use one of
the standard Runge-Kutta schemes to solve it. The field evolution is
much more problematic. It involves the solution of the Cauchy problem for a
non-linear hyperbolic partial differential equation with discontinuous
initial data. In the next chapter we will outline an exact numerical 
integration scheme for the field equation. Here we will describe an 
approximated one.

The idea is to use the auxiliary functions $X(r,t)$ and $Y(r,t)$ introduced
in section \ref{initial-condition}. We make the following approximation:
in the timestep $dt$ we evolve the field according to equation (\ref{afe})
where we consider the source term $F$ as constant in time
\footnote{Note that this is an approximation even within the mean-field 
scheme since in its definition $F$ contains the field itself.}. 
Under this approximation, given at time $t_o$, $X(r,t_o)=X_o(r)$, 
and $Y(r,t_o)=Y_o(r)$ the solutions for $X$ and $Y$ at later times are,
\bq \nonumber
&& X(r,t)=X_o(r+\Delta t)-F\:\mbox{st}[\rp-\Delta t,\rp](r)+F\:
\mbox{st}[-\rp-\Delta t,-\rp](r)~~,\\ \label{asol}
&& Y(r,t)=Y_o(r-\Delta t)+F\:\mbox{st}[\rp,\rp+\Delta t](r)-F\:
\mbox{st}[-\rp,-\rp+\Delta t](r)~~,
\eq
where $\mbox{st}[a,b](r)=H(x-a)-H(x-b)$ with $H$ the Heaviside function,
$\Delta t=t-t_o$, and we added an image soruce at $r=-r_p(t)$
\footnote{See appendix (\ref{mi}) for a justification of our use of the 
images method in the solution of this particular field equation.}
in order to ensure the finiteness of the field at the origin at
all times, which requires,
\bq
X(0,t)=Y(0,t)~~~~~~~\forall t~~.
\eq
We then reconstruct the gravitational field as follows,
\bq
\Phi(r,t)=\frac{1}{r}\int_0^r[X(r,t)+Y(r,t)]\:dr~~.
\eq

In our code we tabulate the field using a uniform grid in $r$ and 
we choose $dr=dt$.
We need infact, to make sure that in using the solutions 
(\ref{asol}), the source terms fall upon the translated functions 
less frequently as possible. Those events are purely due to 
the mean field scheme, which require that we move the particle over 
a fixed field. When $dr=dt$ they occur only when the particle hits 
a grid point at a given timestep.

\chapter[Exact Numerical Integration]{Exact Numerical Integration}

Here we describe the scheme we use to solve exactly the scalar field 
equation (\ref{fe}) coupled to the particle equations (\ref{pe}) in 
spherical symmetry, within the mean-field approximation described 
in secction \ref{cha}.

\section[Characteristics method]{Characteristics method}
\label{characteristics}

In order to make to make the characteristics approximation 
an exact integration we need to replace the solution (\ref{asol}) 
with,
\bq \nonumber
X(r,t)&=&X_o(r+\Delta t)-F(r_p,t+(r-r_p)) \mbox{st}[r_p-\Delta
t,r_p]+\\ \nonumber
&&F(r_p,t+(r+r_p)) \mbox{st}[-r_p-\Delta t,-r_p]\\ \label{bsol}
Y(r,t)&=&Y_o(r+\Delta t)+F(r_p,t-(r-r_p)) \mbox{st}[r_p-\Delta
t,r_p]-\\ \nonumber
&&F(r_p,t+(r-r_p)) \mbox{st}[-r_p,-r_p+\Delta t]
\eq
In our numerical integration we have always used the field time-step
$\Delta t$, equal to the particle time-step $dt$, equal to the grid spacing
$dr$. In this case there is no difference in using equations
(\ref{bsol}) or (\ref{asol}). If we want to use $\Delta t =ndt$ with
$n=2,3,\ldots$ then the more general solution (\ref{bsol}) should be
used and solved by iteration.

\section[High resolution method]{High resolution method}
\label{high}

A more rigorous method when computing discontinuous solutions of the
wave equation can be found among the flux-limiter methods described in
chapter 16 of Randall J. LeVeque ``Numerical Methods for Conservation
Laws''.
Here we will describe the one employing the ``Van Leer'' smoother limiter
function.

This method is second order accurate on smooth parts of the field and
yet gives a well resolved, nonoscillatory discontinuity at the shell
surface (by increasing the amount of numerical dissipation in its
neighborhood). The method has the total variation diminishing property
provided that the Courant, Friedrichs, and Lewy (CFL) condition is
satisfied and consequently it is mononotonicity preserving.

We will first state the method for a general linear hyperbolic system
of partial differential equations and later specialize it to our
nonlinear field equation.

Consider the time-dependent Cauchy problem in one space dimension,
\bq \nonumber
&&u_{,t}+Au_{,x}=0 ~~,~~~~-\infty<x<\infty~~,~~~~t\ge 0\\ \nonumber
&&u(x,0)=u_o(x)~~,
\eq
where $u\in R^m$ and $A$ is an $n\times n$ matrix. The system is
called hyperbolic when $A$ is diagonalizable with real eigenvalues, so
that we can decompose $A=R\Lambda R^{-1}$, where $\Lambda=\mbox{diag}
(\lambda_1,\lambda_2\ldots,\lambda_m)$ is the diagonal matrix of eigenvalues
and $R=[r_1|r_2|\cdots|r_m]$ is the matrix of right eigenvectors of $A$.
Discretize time as $t_n=n dt$ and space as $x_j=j dr$. The finite
difference method we want to describe produces approximations
$U^n_j\in R^m$ to the solution $u(x_j,t_n)=u_j^n$ at the discrete grid
points. The method is written in conservative form as follows,
\bq \label{ds0}
&&U_j^{n+1}=U_j^n - \frac{dt}{dr}(FL^n_j-FL^n_{j-1})~~,\\
&&FL_j=FLl_j+FLh_j~~,\\
&&FLl_j=\frac{1}{2}A(U_j+U_{j+1})-\frac{1}{2}|A|(U_{j+1}-U_j)~~,\\
&&|A|=R(\Lambda^+-\Lambda^-)R^{-1},~\Lambda^\pm=\mbox{diag}(
\lambda_1^\pm,\ldots,\lambda_m^\pm),~\lambda_p^\pm=
\frac{\max}{\min}(\lambda_p,0)~~,\\
&&FLh_j=\frac{1}{2}\sum_{p=1}^m\phi({\theta_p}_j)(\mbox{sgn}(\nu_p)-\nu_p)
\lambda_p {\alpha_p}_j r_p~~,\\
&&\nu_p=\lambda_p \frac{dt}{dr}~~,\\
&&\alpha_j=R^{-1}(U_{j+1}-U_j)~~,\\
&&\phi(\theta)=\frac{|\theta|+\theta}{1+|\theta|}~~,~~~~\mbox{``Van
Leer'' smoother limiter function}\\
&&{\theta_p}_j=\frac{{\alpha_p}_{j^\prime}}{{\alpha_p}_j}~~,~~~~j^
\prime=j-\mbox{sgn}{\nu_p}   
\eq
$FLh$ is the high order (Lax-Wendroff) flux acting on the smooth
portions of the solution (where $\theta$ is near to 1) while $FLl$
is the low order (first order upwind) flux acting in the vicinity of a
discontinuity (where $\theta$ is far from 1). The CFL condition is,
\bq 
\left|\frac{\lambda_p dt}{dr}\right|\le 1~~,~~~~\forall p~~.
\eq 

Our field equation is a wave equation with a nonlinear source term. 
It can be rewritten as follows,
\bq
&&u_{,t}+Au_{,x}=b~~,\\
&&u=\left(
\begin{array}{c}
u1\\
u2
\end{array}
\right)=\left(
\begin{array}{c}
\Psi_{,x}\\
\Psi_{,t}
\end{array}
\right)~~,~~~~\Psi(x,t)=x\Phi(x,t)~~,\\
&&A=\left(
\begin{array}{cc}
0 & -1\\
-1 & 0
\end{array}
\right)~~,~~
|A|=\left(
\begin{array}{cc}
1 & 0\\
0 & 1
\end{array}
\right)~~,\\
&&\Lambda=\left(
\begin{array}{cc}
-1 & 0\\
0 & 1
\end{array}
\right)~~,~~
R=\left(
\begin{array}{cc}
1 & 1\\
1 & -1
\end{array}
\right)~~,~~
R^{-1}=\left(
\begin{array}{cc}
1/2 & 1/2\\
1/2 & -1/2
\end{array}
\right)~~,\\
&&b=\left(
\begin{array}{c}
0 \\
4\pi G \sigma e^{\frac{1}{x}\int_0^xu1(x^\prime,t)\:dx^\prime}x
[\delta(x-r_p)+\delta(x+r_p)]
\end{array}
\right)~~.\\
\eq
The initial condition is,
\bq
u_o(x)=\left(
\begin{array}{c}
-a_i\mbox{st}[-r_i,r_i](x) \\
0 
\end{array}
\right)~~.
\eq
If we call $r_{max}=j_{max}dr$ the maximum extent of our grid, 
the outgoing wave boudary conditions are,
\bq \label{owbc}
u1(x>x_{max},t)+ u2(x>x_{max},t)=0~~,~~~~\forall t~~,\\ \nonumber
u1(x<-x_{max},t)- u2(x<-x_{max},t)=0~~,~~~~\forall t~~.
\eq
Immagine that we have approximated the true solution of the field
equation up to the n-th time slice (i.e. we know $U^n_j$ for
$j=-j_{max},\ldots,-1,0,1,\ldots,j_{max}$). The difference scheme,
\bq
U1^{n+1}_j=f(U1_{j-1}^n,U1_j^n,U1_{j+1}^n,U2_{j-1}^n,U2_j^n,U2_{j+1}^n)~~,\\
U2^{n+1}_j=g(U1_{j-1}^n,U1_j^n,U1_{j+1}^n,U2_{j-1}^n,U2_j^n,U2_{j+1}^n)~~,
\eq
when evaluated at $j_{max}$ becomes a system of 2 equations in 2
unknowns $U1^{n+1}_{j_{max}}$ and $U1^n_{j_{max}+1}$,
\bq
U1^{n+1}_{j_{max}}=f(U1_{j_{max}-1}^n,U1_{j_{max}}^n,U1_{j_{max}+1}^n,
U2_{j_{max}-1}^n)~~,\\ 
U1^{n+1}_{j_{max}}=-g(U1_{j_{max}-1}^n,U1_{j_{max}}^n,U1_{j_{max}+1}^n,
U2_{j_{max}-1}^n)~~,
\eq
allowing the closure of the difference scheme. A consistency check
would be to monitor the constraint,
\bq
U2^n_0=0~~,~~~~\forall n~~.
\eq 

The difference scheme to be used for the field equation follows from
equation (\ref{ds0}),
\bq
U_j^{n+1}=U_j^n - \frac{dt}{dr}(FL^n_j-FL^n_{j-1}) + dt\: B^n_j~~,
\eq 
where $B$ is the approximation to the source term $b$,
\bq \label{sourceB}
B^n_j=\left(
\begin{array}{c}
0\\
4\pi G \sigma(t_n) e^{\frac{1}{x_j}\int_0^{x_j}u1(x,t)\:dx}x_j
\frac{W(r_p(t_n)-x_j)-W(-r_p(t_n)-x_j)}{dr}
\end{array}
\right)~~.
\eq
In equation (\ref{sourceB}) we have approximated the delta functions
using a triangular shaped cloud scheme, which in one dimension employs
3 mesh points and has an assignment-interpolation function $W$ which
is continuous in value and first derivative. Mass is assigned from the
particle at $r_p$ to the 3 mesh points nearest to it,
\bq
W(x)=\left\{
\begin{array}{ll}
\frac{3}{4}-\left(\frac{x}{dr}\right)^2 & |x|\le\frac{dr}{2}\\
\frac{1}{2}\left(\frac{3}{2}-\frac{|x|}{dr}\right)^2 & \frac{dr}{2}
\le|x|\le\frac{3dr}{2}\\
0 & \mbox{otherwise}
\end{array}
\right.~~.
\eq

\chapter[Numerical results]{Numerical results}

When analyzing our numerical results, we will adopt gravitational
units where $G=c=1$. In this chapter we report the results obtained
with the characteristic approximation code (see section \ref{code}).
We will refer to this results as the ``exact integration'' results.

\section[Relaxation to virial equilibrium]{Relaxation to virial equilibrium}

\begin{figure}[hbt] 
\centerline{\includegraphics[height=7.5truecm,width=9truecm]{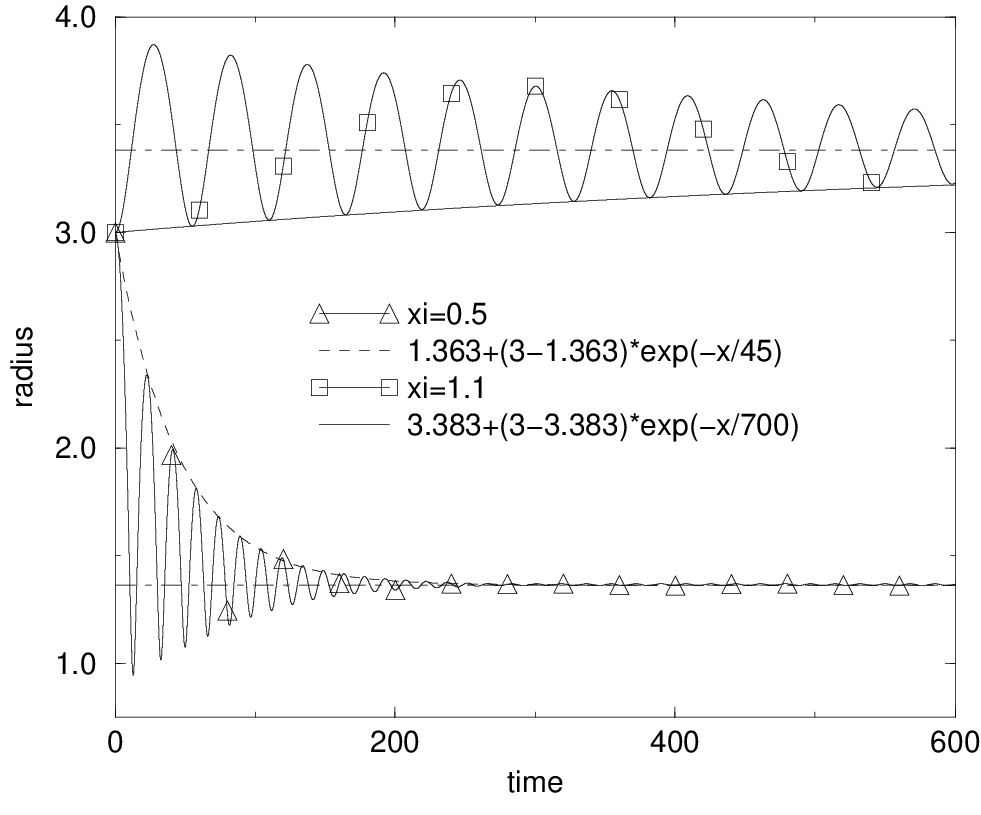}}
\caption{Shows the relaxation to the virial equilibrium state for an 
$\alpha=3$ shell with two different values of $\xi$. In both cases
the decay is fitted well by an exponential.
\label{Nbehave}}
\end{figure}

When trying to reproduce the expected behaviour described in 
figure \ref{behave} we got figure \ref{Nbehave}.

\begin{figure}[hbt]
\centerline{\includegraphics[height=8truecm,width=9truecm]{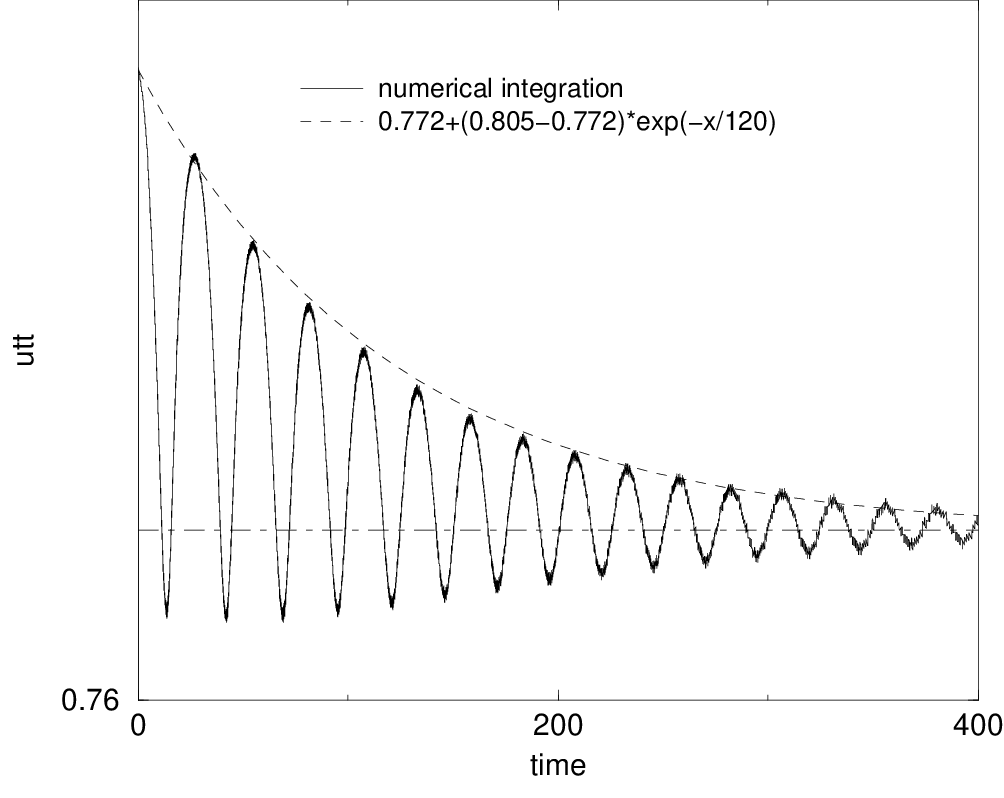}}
\caption{For the case $\alpha=3$, $\xi=0.7$ shows the particle energy 
$\utt$ versus time. The decay to the equilibrium value is well fitted 
by an exponential. 
\label{Nutt}}
\end{figure}

When trying to reproduce the expected behaviour described in 
figure \ref{utt} we got figure \ref{Nutt}.

\section[Comparison with the analytic method]
{Comparison with the analytic method}

We compare the numerical integration in the linear and nonlinear
regimes with the analytic Newtonian solution.
\vspace{3cm}
\begin{figure}[hbt] 
\centerline{\includegraphics[height=6cm]{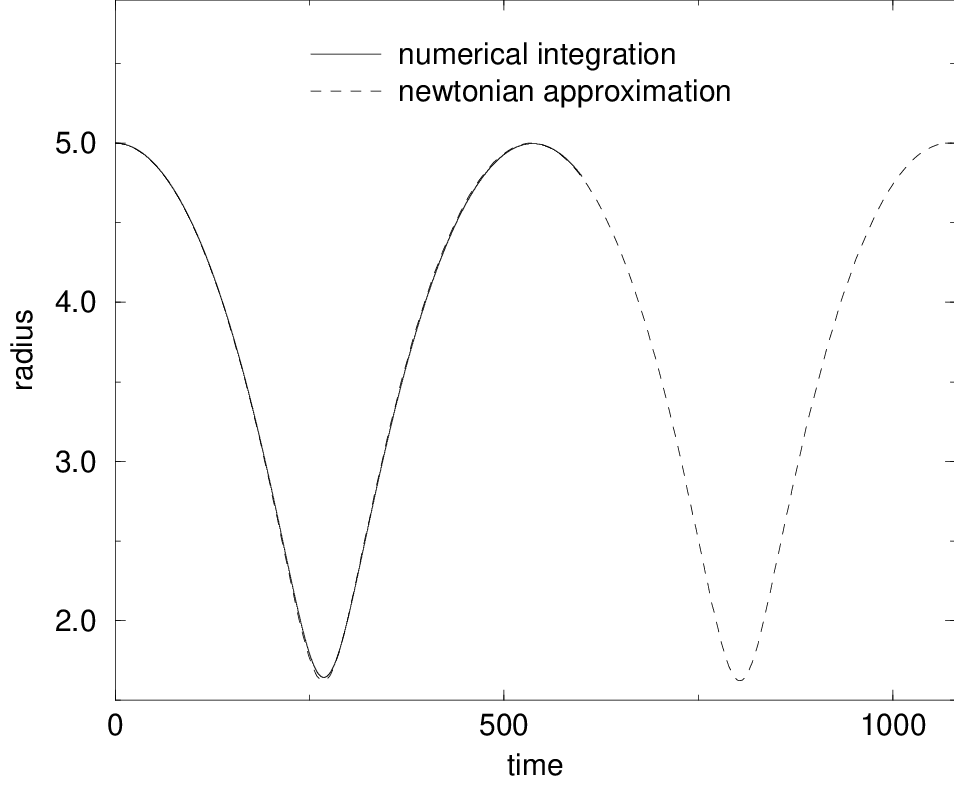}
\includegraphics[height=6cm]{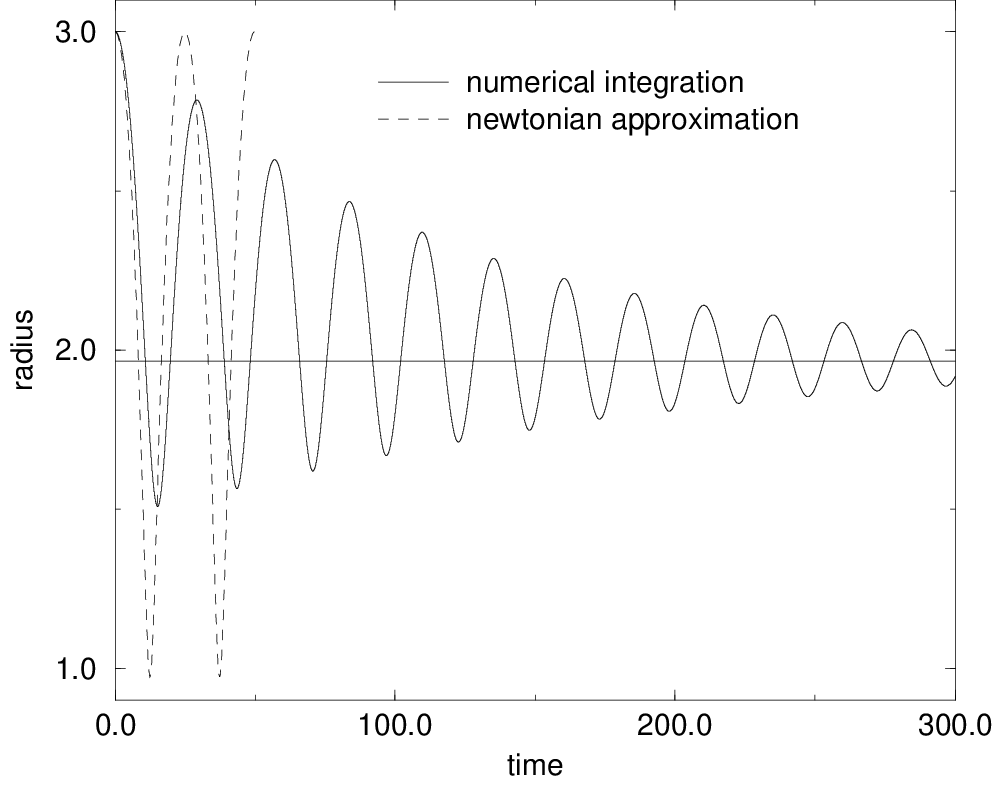}}
\caption{Compares the numerical integration with the analytic Newtonian
approximation. To the left the quasi-Newtonian $\alpha=500$, $\xi=0.7$ 
shell is shown. The predicted equilibrium radius is at $r_e=2.4662896500$.
To the right the $\alpha=3$, $\xi=0.7$ shell is shown. The predicted 
equilibrium radius is at $r_e=1.9657627134$.
\label{a5}}
\end{figure}
\begin{figure}[hbt] 
\centerline{\includegraphics[height=6.5truecm,width=8truecm]{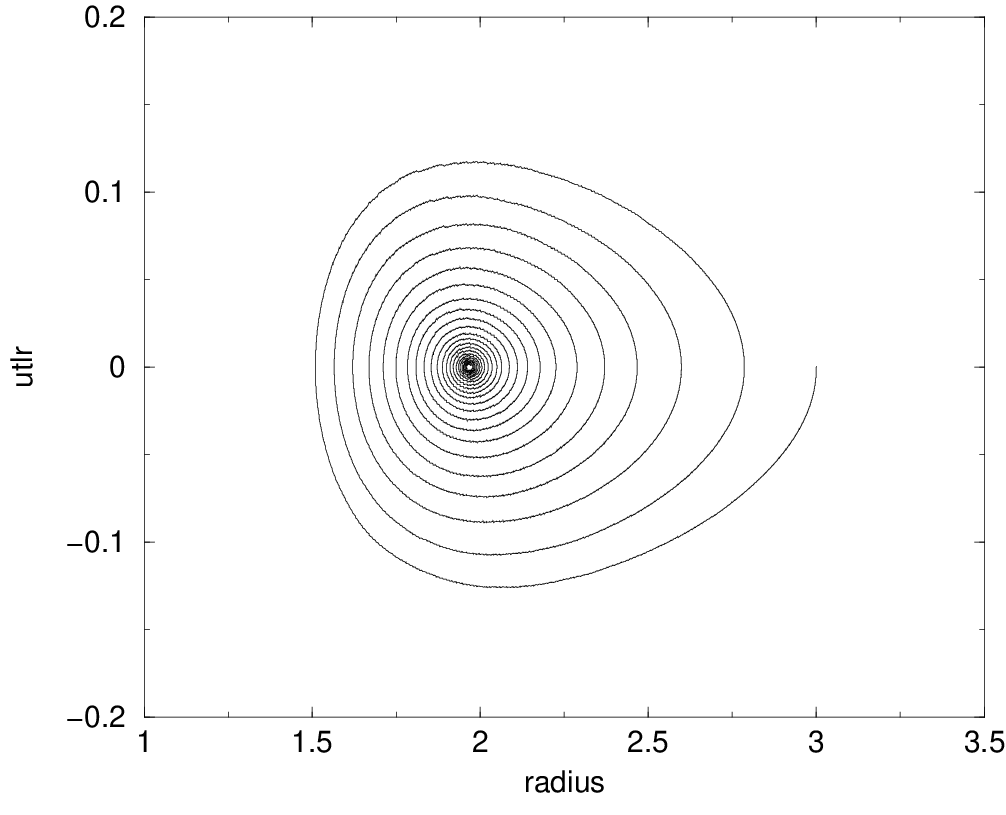}}
\caption{Shows $\utlr$ as a function of the shell radius for the 
case $\alpha=3$, $\xi=0.7$.
\label{phsp1}}
\end{figure}
\begin{figure}[hbt] 
\centerline{\includegraphics[height=6.5truecm,width=8truecm]{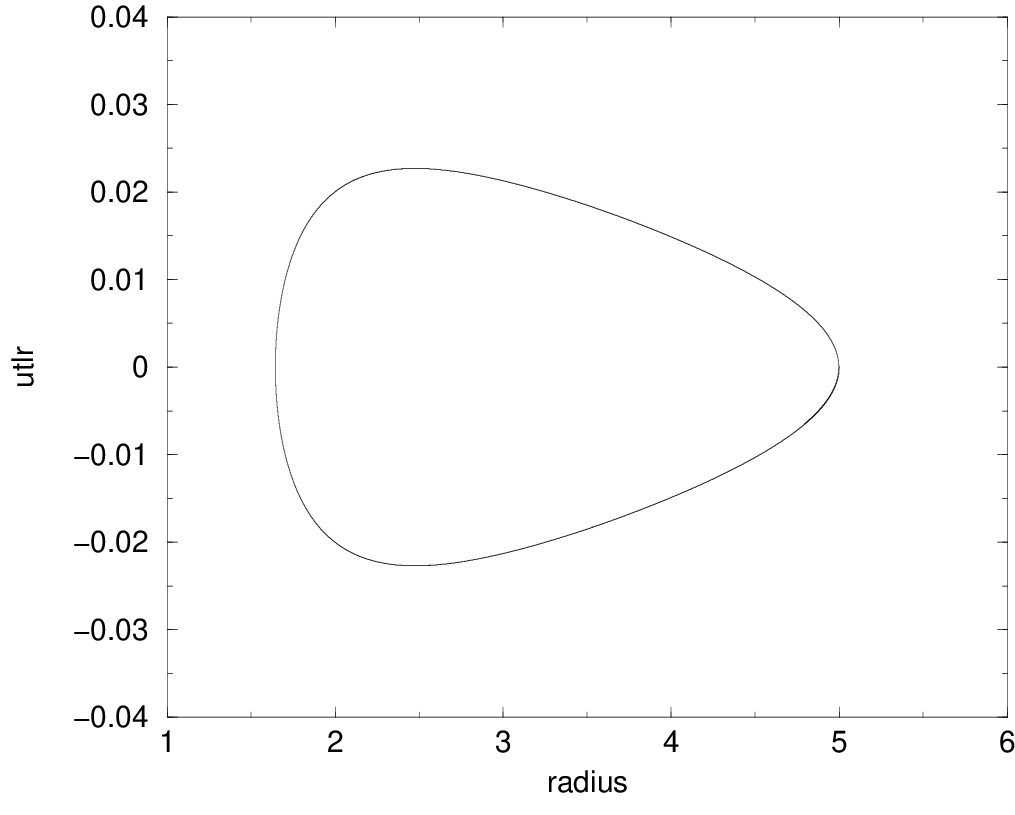}}
\caption{Same as figure \ref{phsp1} for the case $\alpha=500$, $\xi=0.7$. 
\label{phsp3}}
\end{figure}

\section[Monopole radiation]{Monopole radiation}

\begin{figure}[hbt] 
\centerline{\includegraphics[height=7truecm,width=8.5truecm]{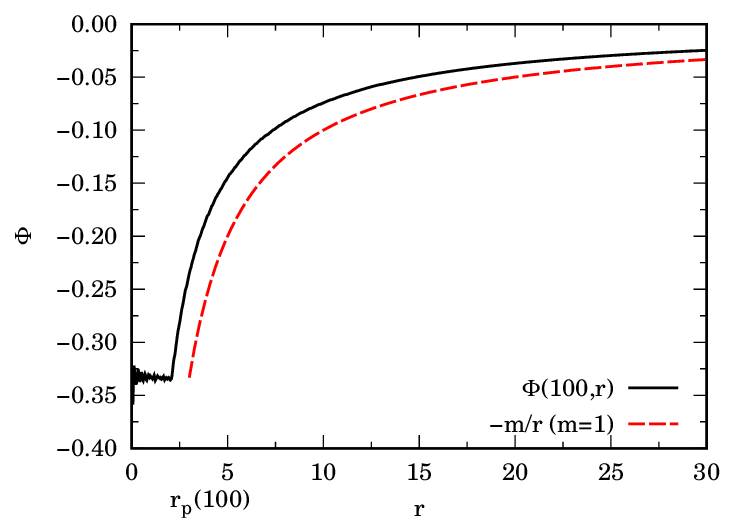}}
\caption{For the case $\alpha=3$, $\xi=0.7$, shows a snapshot at $t=100$
of the field $\Phi(100,t)$, and the zeroth order radiation part $-m/r$.
\label{Nwave}}
\end{figure}

When trying to reproduce the expected behaviour described in 
figure \ref{wave} we got figure \ref{Nwave}.

\section[Quasistatic approximation]{Quasistatic approximation}

\begin{figure}[hbt]
\centerline{\includegraphics[height=6.5truecm,width=8truecm]{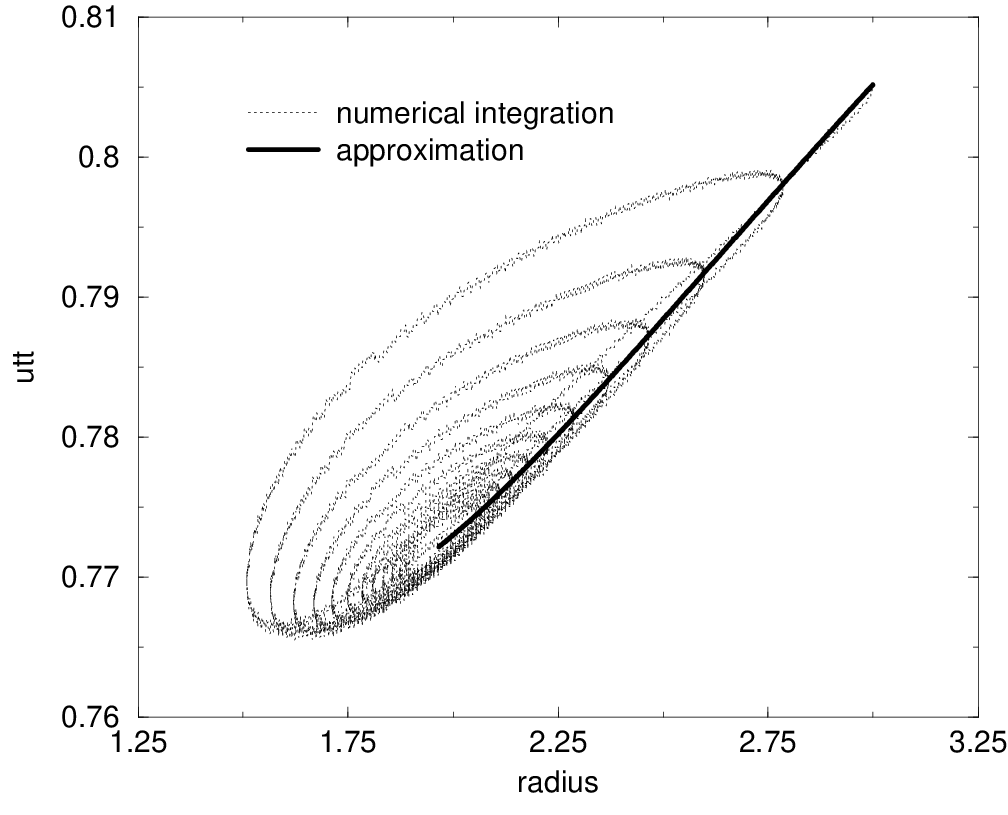}}
\caption{For the $\alpha=3$, $\xi=0.7$ case, shows the $\utt$ as a function 
of the shell radius for the numerical integration. The solid line was 
derived using the analytic expression (\ref{uttqs}). We see that it 
approximates well the values for the energy at the turning points.
\label{Nphsp2}}
\end{figure}

\begin{figure}[hbt]
\centerline{\includegraphics[height=6.5truecm,width=8truecm]{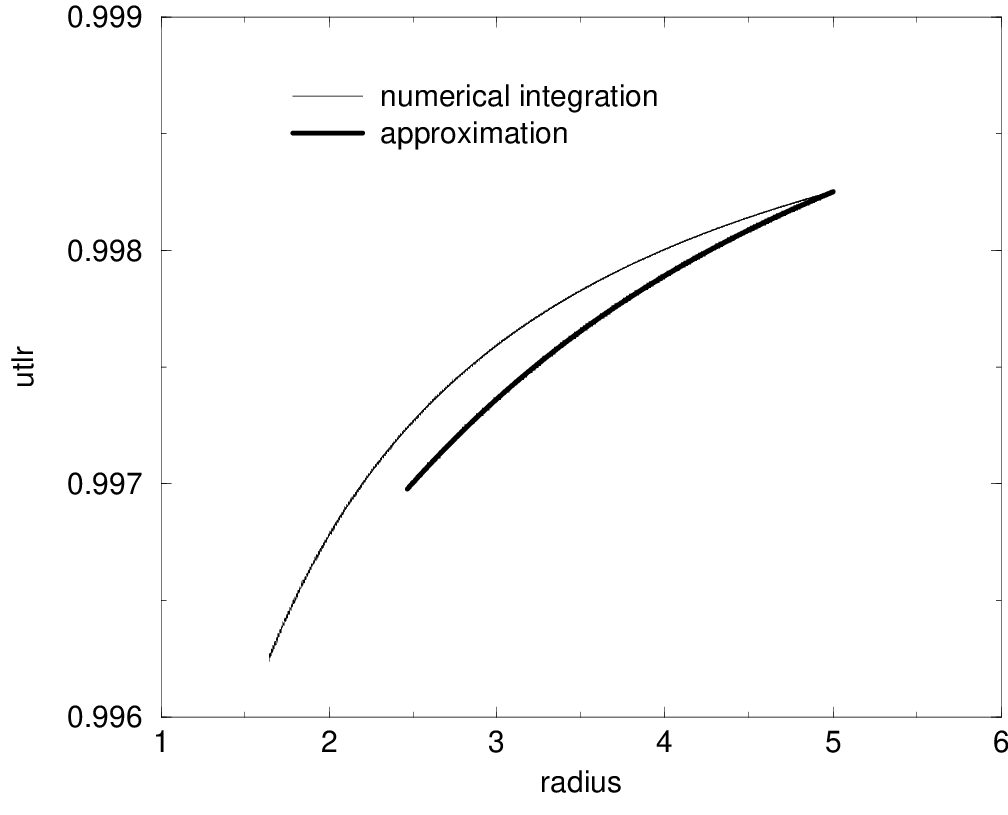}}
\caption{Same as figure \ref{phsp2} but for the $\alpha=500$, $\xi=0.7$ 
case.
\label{Nphsp4}}
\end{figure}

When trying to reproduce the expected behaviour described in 
figure \ref{phsp2} we got figures \ref{Nphsp2} and \ref{Nphsp4}.

\chapter[Conclusions]{Conclusions}

Some future developments to the present work may be:
\begin{description}
\item[0] Correct the characteristics approximation as outlined in section
ref{characteristics} 
\item[1] Integrate the equations (\ref{fe}) and (\ref{particle}) using
the finite-difference scheme for the evolution of the field 
described in section \ref{high}.
\item[2] Extend the one particle problem to a many particle one, and 
check how the quasistatic approximation performs there.
\item[3] Go on to solve more realistic gravitational field theories, and 
look for quasistatic approximations.
\end{description}

\clearpage
\appendix
\chapter[Equations of motion in spherical symmetry]{Equations of motion in spherical symmetry}
\label{app:0}

In this Appenix we will derive the equations of motion in spherical symmetry presented in Eq. 
(\ref{pe}). Spherical symmetry implies that $\Phi_{,\theta}=\Phi_{,\phi}=0$.
In spherical coordinates $g_{\alpha\beta}=\mbox{diag}(-1,1,r^2,r^2\sin^2\theta)$ which on the 
plane $\theta=\pi/2$ becomes $g_{\alpha\beta}=\mbox{diag}(-1,1,r^2,r^2)$.

From the definitions $u^\alpha=dz^\alpha/d\tau$ and $\tilde{u}^\alpha=e^{\Phi}u^\alpha$ and 
$ds^2=g_{\alpha\beta}dz^\alpha dz^\beta$ follows immediately Eq. (\ref{penergy}) where
$u^r=u_r$ and $u_\phi=r^2 u^\phi=r^2u^0\dot{\phi}$, the dot is a derivative with respect to time 
$t=z^0$ as usual, and $u^0=dt/d\tau=\gamma$ the Lorentz factor.

Note that $Du^\alpha/dt={u^\alpha}_{;0}=\dot{u}^\alpha+u^\mu{\Gamma^\alpha}_{\mu 0}$ where the only 
non-zero Christoffel symbol is ${\Gamma^\phi}_{\phi 0}=\dot{r}/r=\tilde{u}^r/r\tilde{u}^0$.

From the particle equation of motion (\ref{particle}) we find
\bq \label{eq:motion}
u^0 u^\alpha_{;0}+{\Phi_,}^\alpha+u^\alpha(u^r\Phi_{,r}+u^0\Phi_{,0})=0.
\eq

We then find
\bq \nonumber
\dot{\tilde{u}}^\phi &=&e^\Phi \Phi_{,0}u^\phi+e^\Phi\dot{u}^\phi\\
&=&-\frac{\tilde{u}^\phi\tilde{u}^r}{\tilde{u}^0r}-\frac{\tilde{u}^\phi\tilde{u}^r}{\tilde{u}^0}\Phi_{,r},\\ \nonumber
\dot{\tilde{u}}^r &=&e^\Phi \Phi_{,0}u^r+e^\Phi\dot{u}^r\\ \label{eq:utdr}
&=&-\frac{e^{2\Phi}\Phi_{,r}}{\tilde{u}^0}-\frac{\tilde{u}_r^2}{\tilde{u}^0}\Phi_{,r},
\eq

So that since $\tilde{u}_\phi=r^2\tilde{u}^\phi$ we have
\bq \nonumber
\dot{\tilde{u}}_\phi&=&2r\dot{r}\tilde{u}^\phi+r^2\dot{\tilde{u}}^\phi\\ \nonumber
&=&r\frac{\tilde{u}^\phi\tilde{u}^r}{\tilde{u}^0}-r^2\frac{\tilde{u}^\phi\tilde{u}^r}{\tilde{u}^0}
\Phi_{,r}\\
&=&0,
\eq
where in the last equality we used the fact that in the reference frame of the 
particle the scalar gravitational field $\Phi$ is static so that its equation 
(\ref{fe}) on the $\theta=\pi/2$ plane becomes
\bq
\frac{1}{r}\left(r\Phi_{,r}\right)_{,r}=-2\pi\delta^2(\zb), 
\eq
which has solution $\Phi(r)=\ln r$. Then we conclude that the particle `orbital 
angular momentum' $\tilde{u}_\phi$ is conserved.

On the other hand since $\tilde{u}_r=\tilde{u}^r$ we find from Eq. (\ref{eq:utdr})
\bq \nonumber
\dot{\tilde{u}}_r &=&-\frac{e^{2\Phi}\Phi_{,r}}{\tilde{u}^0}-\frac{\tilde{u}_r^2}{\tilde{u}^0}\Phi_{,r}\\ \nonumber
&=&-\frac{e^{2\Phi}\Phi_{,r}}{\tilde{u}^0}+\frac{\tilde{u}_\phi^2}{r^2\tilde{u}^0}\Phi_{,r}\\
&=&-\frac{e^{2\Phi}\Phi_{,r}}{\tilde{u}^0}+\frac{\tilde{u}_\phi^2}{r^3\tilde{u}^0}
\eq
where we used the fact that in the comoving frame $dt=d\tau$ and $(dr)^2+(rd\phi)^2=0$.

\chapter[Energy loss in the static approximation]{Energy loss in the static approximation}

In the weak field slow motion limit, in the wave zone the gravity 
wave amplitude can be written (dropping terms higher than the monopole)
as (see equation (\ref{srad}) in the main text),
\bq \nonumber
r\Phi(r,t)=-G\int d\rpp\: 4\pi\rpp^2[\rho_0(\Phi-\shalf v^2)
+\ssixth \rp^2{\rho_0}_{,tt}]_{t-r}~~,
\eq
where in the static approximation,
\bq
&&\rho_o=\frac{m}{4\pi r_s^2} \delta(\rpp-r_s)~~,\\
&&v^2=(\dot{r}_s)^2+\left(\frac{\utlp}{\utt}\right)^2
\frac{1}{r_s^2}~~,\\
&&\Phi=\left\{
\begin{array}{ll}
a_s/r_s & r\leq r_s\\
a_s/r   & r>r_s
\end{array}
\right.~~,
\eq 
and $a_s=a_s(r_s,\utlr)$ through the jump condition (see equation (\ref{nf}) 
in the main text),
\bq \nonumber
a_s=-\frac{Gm\:e^{\textstyle 2a_s/r_s}}{\sqrt{e^{\textstyle 2a_s/r_s}+\utlr^2
+\utlp^2/r_s^2}}~~.
\eq
Then we can rewrite the wave amplitude as follows,
\bq \nonumber
r\Phi(r,t)&=&-Gm\left\{ \frac{a_s}{r_s}-\frac{(\dot{r}_s)^2}{2}
-\left(\frac{\utlp}{\utt}\right)^2\frac{1}{2r_s^2}+\frac{1}{3}
[(\dot{r}_s)^2+r_s \ddot{r}_s]\right\}_{t-r}~~,\\
&=&-Gm\left\{ \frac{a_s}{r_s}-\left(\frac{\utlp}{\utt}\right)^2
\frac{1}{2r_s^2}-\frac{1}{6}(\dot{r}_s)^2+\frac{1}{3}r_s \ddot{r}_s
\right\}_{t-r}~~.
\eq
Taking the time derivative one gets (both $\utt$ and $\utlp$ are
constants of motion),
\bq \label{a-eloss}
r\Phi_{,t}&=&-Gm\left\{\frac{\dot{a}_s}{r_s}-
\frac{a_s\dot{r}_s}{r_s^2}+\left(\frac{\utlp}{\utt}\right)^2
\frac{\dot{r}_s}{r_s^3}+\frac{1}{3}r_s\dot{\ddot{r}}_s 
\right\}_{t-r}~~.
\eq
In the static approximation,
\bq \label{a-f}
&&\dot{r}_s=\frac{\utlr}{\utt}~~,\\
&&\ddot{r}_s=\frac{\dot{\utlr}}{\utt}=\left(\frac{\utlp}{\utt}\right)^2
\frac{1}{r_s^3}-\frac{e^{2a_s/r_s}}{(\utt)^2}\frac{a_s}{2r_s^2}~~,\\
&&\dot{\ddot{r}}_s=-\left(\frac{\utlp}{\utt}\right)^2
\frac{3\dot{r}_s}{r_s^4}-\frac{e^{2a_s/r_s}}{2(\utt)^2}\left(
\frac{\dot{a}_s}{r_s^2}-\frac{2a_s\dot{r}_s}{r_s^3}\right)-
\frac{e^{2a_s/r_s}}{(\utt)^2}\frac{a_s}{r_s^2}\left(
\frac{\dot{a}_s}{r_s}-\frac{a_s\dot{r}_s}{r_s^2}\right)~~,\\
&&\dot{a}_s=(a_s)_{,\utlr}\utt\ddot{r}_s+(a_s)_{,r_s}\dot{r}_s~~,\\
&&(a_s)_{,\utlr}=-\frac{\utlr\frac{a}{e^{2a_s/r_s}+\utlr^2+\utlp^2/r_s^2}}
{1-\frac{2a_s}{r_s}+\frac{a_s}{r_s}\frac{e^{2a_s/r_s}}
{e^{2a_s/r_s}+\utlr^2+\utlp^2/r_s^2}}~~,\\ \label{a-l}
&&(a_s)_{,r_s}=-\frac{\frac{2a_s^2}{r_s^2}-\frac{a_s^2}{r_s^2}
\frac{e^{2a_s/r_s}}{e^{2a_s/r_s}+\utlr^2+\utlp^2/r_s^2}}
{1-\frac{2a_s}{r_s}+\frac{a_s}{r_s}\frac{e^{2a_s/r_s}}
{e^{2a_s/r_s}+\utlr^2+\utlp^2/r_s^2}}~~.
\eq

Using equations (\ref{a-f})-(\ref{a-l}) into equation (\ref{a-eloss}) 
one can determine numerically the rate of energy loss (\ref{erate}).
This can then be integrated to get the energy emitted by the particle 
in a full revolution around the origin. This calculation can 
be carried out analytically in the Newtonian approximation as shown
in detail in the next section.

\section[Newtonian approximation]{Newtonian approximation}

In the Newtonian approximation we have,
\bq
&&a_s   \rightarrow -Gm~~,\\
&&\frac{\utlp}{\utt} \rightarrow \sqrt{\frac{Gm\xi^2 r_i}{2}}~~,\\
&&\ddot{r}_s \rightarrow -\frac{Gm}{2r_s^2}+\frac{Gm\xi^2 r_i}{2r_s^3}~~.
\eq
Making these substitutions in equation (\ref{a-eloss}) we get equation
(\ref{eloss}) of the main text,
\bq
r\Phi_{,t}=-\frac{4}{3}\frac{(Gm)^2}{r_i}\left[\frac{\dot{x}}{x^2}
\right]_{t-r}~~,
\eq
where $x=r_s/r_i$. So for the rate of energy emission in the wave zone 
we get equation (\ref{Nerate}), which integrated over one orbital 
period gives equation (\ref{deltaE}).

\chapter[Method of Images]{Method of Images}
\label{mi}

We want to justify the use we have made of the images method,
in the solution of the nonlinear field equation (\ref{fe}).

To do that we need to show the equivalence between the two 
following problems. Calling $\Psi(r,t)=r\Phi(r,t)$, with 
$r\in [0,\infty]$, the first problem is our original one,
namely,
\bq
\mbox{problem 1:}\left\{
\begin{array}{ll}
\Psi_{,tt}-\Psi_{,rr}=F(r,\Psi(r,t))\delta(r-r_p)\\
\Psi(r,0)=f(r)&\mbox{i.c.}\\
\Psi_{,t}(r,0)=0&\mbox{i.c.}\\
\Psi(0,t)=0&\mbox{b.c.}\\
\Psi_{,r}(r_m,t)+\Psi_{,t}(r_m,t)=0&\mbox{b.c.}\\
\end{array}
\right.
\eq 
where i.c. stands for initial condition and b.c. for boundary condition.

The second problem is over the whole real axis $x\in[-\infty,\infty]$
and employs two sources, the one at $r_p$, of the first problem, and its
image,
\bq
\mbox{problem 2:}\left\{
\begin{array}{ll}
\Psi_{,tt}-\Psi_{,xx}=F(x,\Psi(x,t))[\delta(x-r_p)+\delta(x+r_p)]\\
\Psi(x,0)=f(x)-f(-x)&\mbox{i.c.}\\
\Psi_{,t}(x,0)=0&\mbox{i.c.}\\
\Psi_{,x}(\pm r_m,t)\pm\Psi_{,t}(\pm r_m,t)=0&\mbox{b.c.}\\
\end{array}
\right.
\eq 

The general solution to problem 1 can be written in integral form as follows,
\bq \label{sol1}
\Psi(r,t)=\frac{1}{2}[f(r+t)+f(r-t)+W_{r_p}(r,t)]
-\frac{1}{2}[r\rightarrow -r]~~,
\eq 
where,
\bq
W_{r_p}(r,t)=\int_0^t d\overline{t}\;F(r_p,\Psi(r_p,t))[
H(r_p-r+(t-\overline{t}))-H(-r_p+r+(t-\overline{t}))]
\eq 
and the last term in equation (\ref{sol1}) was added in order to have
the solution satisfy the boundary condition at $r=0$. The outgoing wave
boundary condition is automatically satisfayed since $r_m$ is intended 
to be at all times to the right of the source, and $f(r)$ is constant 
for $r>r_p(0)$. So there are no ingoing waves passing through $r_m$.

The general solution to problem 2 can be written in integral form as follows,
\bq \label{sol2}
\Psi(r,t)&=&\frac{1}{2}\{[f(x+t)-f(-x-t)]+[f(x-t)-f(-x+t)]\\
&&+W_{r_p}(x,t)+W_{-r_p}(x,t)\}~~,
\eq 

In order for the two problems to have the same solution for $x\ge 0$,
the following condition has to be satisfied,
\bq
W_{r_p}(-x,t)=-W_{-r_p}(x,t)~~.
\eq
This condition is equivalent to, 
\bq
F(r_p,\Psi(r_p,t))=-F(-r_p,\Psi(-r_p,t))=-F(-r_p,-\Psi(r_p,t))~~,
\eq
where in the last equality we used the fact that the field is an odd
function in $x$ at all times. We can easily verify that our field 
equation, where,
\bq
F(r_p,\Psi(r_p,t))=-\frac{Gm}{r_p}\frac{e^{2\Psi(r_p,t)/r_p}}
{\sqrt{e^{2\Psi(r_p,t)/r_p}+\utlr^2+\utlp^2/r_p^2}}~~,
\eq
satisfies such condition.

\chapter[The nonhomogeneous wave equation]{The nonhomogeneous wave equation}

We want to find a solution to the following problem,
\bq
&&\Psi_{,tt}-\Psi_{,xx}=F(x,t)~~,\\
&&\Psi(x,0)=0~~,\\
&&\Psi_{,t}(x,0)=0~~.
\eq
Make the change of variables,
\bq
\xi=x+t~~,\\
\eta=x-t~~.
\eq
The differential equation then becomes,
\bq
\Psi_{,\xi\eta}\left(
\frac{\xi+\eta}{2},\frac{\xi-\eta}{2}\right)=-\frac{1}{4}F\left(
\frac{\xi+\eta}{2},\frac{\xi-\eta}{2}\right)~~.
\eq
Integrating with respect to $\xi$, we have,
\bq
\Psi_{,\eta}\left(
\frac{\xi+\eta}{2},\frac{\xi-\eta}{2}\right)&=&
\left.\Psi_{\eta}\left(
\frac{\xi+\eta}{2},\frac{\xi-\eta}{2}\right)\right]_{\overline{\xi}=\eta}\\
&&+\int_\eta^\xi \Psi_{,\xi\eta}\left(
\frac{\overline{\xi}+\eta}{2},\frac{\overline{\xi}-\eta}{2}\right)
d\overline{\xi}\\
&&=\frac{1}{2}\Psi_{,x}(\eta,0)-
\frac{1}{2}\Psi_{,t}(\eta,0)\\
&&-\frac{1}{4}\int_\eta^\xi F\left(
\frac{\overline{\xi}+\eta}{2},\frac{\overline{\xi}-\eta}{2}\right)
d\overline{\xi}
\eq
We integrate this equation from an arbitrary value of $\eta$ to $\xi$
to find,
\bq
\Psi(\xi,0)-\Psi\left(
\frac{\xi+\eta}{2},\frac{\xi-\eta}{2}\right)&=&
\int_\eta^\xi\left[\frac{1}{2}\Psi_{,x}
(\overline{\eta},0)-
\frac{1}{2}\Psi_{,t}(\overline{\eta},0)\right]
d\overline{\eta}\\
&&-\frac{1}{4}\int_\eta^\xi\int_{\overline{\eta}}^\xi
F\left(
\frac{\overline{\xi}+\overline{\eta}}{2},\frac{\overline{\xi}-\overline{\eta}}{2}\right)
d\overline{\xi}d\overline{\eta}~~.
\eq
In the first integral we note that,
\bq
\int_\eta^\xi\Psi_{,x}(\overline{\eta},0)
d\overline{\eta}=\Psi(\xi,0)-\Psi(\eta,0)~~.
\eq
In the second integral we let,
\bq
\overline{\eta}=\overline{x}-\overline{t}~~,\\
\overline{\xi}=\overline{x}+\overline{t}~~.
\eq
The domain of integration $\eta\le \overline{\eta}\le \overline{\xi}\le\xi$
becomes
\bq
\eta\le \overline{x}-\overline{t}\le \overline{x}+\overline{t}\le \xi~~,
\eq
or
\bq
\eta+\overline{t}\le\overline{x}\le\xi-\overline{t}~~,~~~~
0\le\overline{t}\le\frac{1}{2}(\xi-\eta)~~.
\eq
The jacobian determinant of the transformation from $(\overline{\xi},
\overline{\eta})$ to $(\overline{x},\overline{t})$ is 2. Therefore
\bq
\frac{1}{4}\int_\eta^\xi\int_{\overline{\eta}}^\xi
F\left(
\frac{\overline{\xi}+\overline{\eta}}{2},\frac{\overline{\xi}-
\overline{\eta}}{2}\right)
d\overline{\xi}d\overline{\eta}=
\frac{1}{2}\int_0^{(\xi-\eta)/2}\int_{\eta+\overline{t}}^{\xi-\overline{t}}
F(\overline{x},\overline{t})d\overline{x}d\overline{t}~~.
\eq
Making these substitutions and transposing, we find
\bq
\Psi\left(
\frac{\xi+\eta}{2},\frac{\xi-\eta}{2}\right)&=&
\frac{1}{2}[\Psi(\xi,0)+\Psi(\eta,0)]+\frac{1}{2}\int_\eta^\xi
\Psi_{,t}(\overline{x},0)d\overline{x}\\
&&+\frac{1}{2}\int_0^{(\xi-\eta)/2}\int_{\eta+\overline{t}}^{\xi-\overline{t}}
F(\overline{x},\overline{t})d\overline{x}d\overline{t}~~.
\eq
We recall that $\xi=x+t$ and $\eta=x-t$. We use the initial conditions to 
obtain the solution formula
\bq
\Psi(x,t)=\frac{1}{2}\int_0^t\int_{x-(t-\overline{t})}^{x+(t-\overline{t})}
F(\overline{x},\overline{t})d\overline{x}d\overline{t}~~.
\eq 
\twocolumn[]
\chapter[The code]{The code}
\label{code}
This is the code used for the exact numerical integration.
{\tiny
\begin{verbatim}
ccccccccccccccccccccccccccccccccccccccccccccccccccccccccccccccc
c                   1 SHELL CLUSTER                                
c dimt = number of timesteps in the integration
c dimg = dimension of the uniform r grid
c 
c INPUT r0=shell radius
c       mr=shell rest mass
c       xi=up/up(circular)
c       dt=time-step
c       rot=dr/dt       dr=grid spacing       
c OUTPUT              
c       fort.8 = (t,rp)
c       fort.9 = (rp,utt,utlr)
c       erp=equilibrium radius
ccccccccccccccccccccccccccccccccccccccccccccccccccccccccccccccc
      implicit none
      include  'cluster.p'
c INPUT
      real*8  r0,mr,xi
      real*8  dt
      integer rot
c OUTPUT
      real*8  erp,rp,utt,utlr 
c INTERNAL
      real*8  phirp,e2p,fpl,fpr,phiprp,e2ppr,st
      real*8  a,ea,dr
      real*8  ut,up,utlp,am2
      real*8  xx(0:imax),yy(-imax:imax)
      real*8  rg(-imax:imax),phi(0:imax)
      integer i,tsteps,jp,dimg,dimt
      parameter(dimg=1000)
      parameter(dimt=60000)
c ========================INPUT DATA============================
      call in(rp,mr,xi,dr,rot,dt)
      r0=rp
c =====================INITIAL CONDITION========================
      tsteps=0
c -------------uniform grid in r (spacing dr)-------------------
      do i=-dimg,dimg
         rg(i)=dble(i)*dr
      enddo
c ----------particle--------------------------------------------
c tangential orbit (utlr=0)
      utlr=0.d0
c find angular velocity for the circular orbit at rp
      call phi1(mr,rp,a,up)
c angular momentum for circular orbits (in a time
c independent field) is:
c utlp(circ)=ulp(circ)*exp(phi)=up(circ)*r*r*exp(phi)      
c set utlp=xi*utlp(circ) = constant of motion
      utlp=xi*rp**2*exp(a/rp)*up
      am2 =utlp**2
      up=utlp/(rp**2*exp(a/rp))
      ut=sqrt(1.d0+(rp*up)**2)
c initial source term
      st=exp(a/rp)*mr/(2.d0*rp*ut)
c ----------field-----------------------------------------------
c xx(r,0)=yy(r,0)=(r*phi(r,0)),r
c phi(r,0)=a/rp    r <= rp
c phi(r,0)=a/r     r >  rp
      jp=nint(rp/dr)
c real space       r >= 0
      do i=0,jp-1
         xx(i)=.5d0*a/rp
         yy(i)=.5d0*a/rp
         phi(i)=a/rp
      enddo
      do i=jp,dimg
         xx(i)=0.d0
         yy(i)=0.d0
         phi(i)=a/rg(i)
      enddo
c immaginary space r < 0
      do i=-dimg,-jp-1
         yy(i)=0.d0
      enddo
      do i=-jp,-1
         yy(i)=.5d0*a/rp
      enddo
c =====================NEXT TIMESTEP============================
 100  tsteps=tsteps+1
      if(mod(tsteps,rot).ne.0) goto 15

c ----------evolve field----------------------------------------
c reinterpolate phi(rp) to find new source term
      jp   =nint(rp/dr)
      phirp=phi(jp)
      e2p  = exp(2.d0*phirp)
      utt  = sqrt(e2p+utlr**2+(utlp/rp)**2)
c the new source term is
      st   = .5d0*e2p*mr/(rp*utt)
c  evolve the field
      call evphi(dimg,dr,rg,st,rp,xx,yy,phi)

c ----------evolve particle-------------------------------------
c find e2p=exp(2*phi(rp)) 
 15   jp    = nint(rp/dr)
      phirp = phi(jp)
      e2p   = exp(2.d0*phirp)
c find e2ppr=e2p*(phi,r(rp-)+phi,r(rp+))/2
      fpl   = (xx(jp-1)+yy(jp-1)-phirp)/rp
      fpr   = (xx(jp+1)+yy(jp+1)-phirp)/rp
      phiprp= (fpl+fpr)*.5d0 
      e2ppr = e2p*phiprp
c evolve the particle with 4-th order Runge-Kutta         
      call runge4(am2,e2p,e2ppr,dt,rp,utlr,rp,utlr) 

c write fort.8 :[t,r(t)] and fort.9 :[t,utt(t),utlr(t)]
      write(8,*) tsteps*dt,rp,phirp
      write(9,*) rp,utt,utlr
      if(tsteps.eq.dimt) goto 200
      goto 100

c estimate the final equilibrium radius erp
 200  call eqrp1(utlp,mr,erp,ea)
c write output
      call out(mr,r0,erp,xi,dt,dr,dimt,dimg)

      stop
      end
\end{verbatim}
\begin{verbatim}
cccccccccccccccccccccccccccccccccccccccccccccccccccccccccccccc
c  Read initial data
c INPUT
c  rp    = initial shell radius
c  mr    = shell rest mass
c  xi    = ratio up/up(circular)
c  dt    = time-step
c  rot   = dr/dt(>=1 Courant stability condition)
c
c OUTPUT(all above +)
c  dr    = grid spacing
cccccccccccccccccccccccccccccccccccccccccccccccccccccccccccccccc
      subroutine in(rp,mr,xi,dr,rot,dt)
      implicit none
      real*8 rp,mr,xi,dr,dt
      integer rot

      write(*,*) 'initial radius rp'
      read(*,*) rp
      write(*,*) 'rest mass mr'
      read(*,*) mr
      write(*,*) 'ratio utlp/utlp(circular)'
      read(*,*) xi 
      write(*,*) 'time-step dt'
      read(*,*) dt
      write(*,*) 'ratio dr/dt=[integer>=1]'
      read(*,*) rot
      dr=dt*dble(rot)

      return
      end
\end{verbatim}
\begin{verbatim}
cccccccccccccccccccccccccccccccccccccccccccccccccccccc
c  4-th order runge-kutta
c  advances to the next time step (h) the equations
c  dr/dt=f(r,u)
c  du/dt=g(r.u)
c  where u=utlr (u tilde-low-r)
c  f(r,u)=u/utu0
c  g(r,u)=utlp**2/(utu0*r**3)-exp(2*phi)*phi,r/utu0
c  utu0=sqrt(exp(2*phi)+u**2+(utlp/r)**2)
c  and phi   = potential at r,u
c      phi,r = d(phi)/dr at r,u
c INPUT  am2(=utlp**2 angular momentum squared),
c        e2p(=exp(2*phi)),e2ppr(=exp(2*phi)*phi,r),
c        h(=time step),ri,ui(=initial values for r,u)
c OUTPUT r,u  (=final   values for r,u)
cccccccccccccccccccccccccccccccccccccccccccccccccccccc
      subroutine runge4(am2,e2p,e2ppr,h,ri,ui,r,u)
      implicit none
c INPUT
      real*8 am2,e2p,e2ppr,h,ri,ui
c OUTPUT
      real*8 u,r
c INTERNAL
      real*8 f,g,u0
      real*8 k1,k2,k3,k4,l1,l2,l3,l4
      real*8 k1o2,k2o2,l1o2,l2o2
      real*8 inv6
      parameter(inv6=1/6.d0)
      u0(r,u)= sqrt(e2p+u**2+am2/r**2)
      f(r,u) = u/u0(r,u)
      g(r,u) = am2/(u0(r,u)*r**3)-e2ppr/u0(r,u)

      k1  = h*f(ri,ui)
      l1  = h*g(ri,ui)
      k1o2= .5d0*k1
      l1o2= .5d0*l1
      k2  = h*f(ri+k1o2,ui+l1o2)
      l2  = h*g(ri+k1o2,ui+l1o2)
      k2o2= .5d0*k2
      l2o2= .5d0*l2
      k3  = h*f(ri+k2o2,ui+l2o2)
      l3  = h*g(ri+k2o2,ui+l2o2)
      k4  = h*f(ri+k3,ui+l3)
      l4  = h*g(ri+k3,ui+l3)

      r   = ri+inv6*(k1+2.d0*(k2+k3)+k4)
      u   = ui+inv6*(l1+2.d0*(l2+l3)+l4)

      return
      end
\end{verbatim}
\begin{verbatim}
cccccccccccccccccccccccccccccccccccccccccccccccccccccccccccc
c  Integrate the 1D wave equation with a delta   
c  function as the source
c      -(r*phi),tt+(r*phi),rr=2*st*delta(r-rp)
c  rewritten as
c      xx,t=xx,r+st*delta(r-rp)
c      yy,t=-yy,r-st*delta(r-rp)
c  where
c      xx=(v+w)/2
c      yy=(v-w)/2
c  and   
c      v=(r*phi),r
c      w=(r*phi),t
c  add an image to ensure finiteness of phi(0,t) forall t
c      xx(0,t)=yy(0,t)  at all times
c
c INPUT  dr (= grid spacing), rg(-imax:imax) (= grid), 
c        st (= source term),  rp (= shell radius), 
c        xxo(0:imax), yyo(-imax:imax) (= old "field")
c
c OUTPUT xxo(0:imax), yyo(-imax:imax), phi(0:imax) (=field)          
cccccccccccccccccccccccccccccccccccccccccccccccccccccccccccc
      subroutine evphi(dimg,dr,rg,st,rp,xxo,yyo,phi)
      implicit none
      include 'cluster.p'
c INPUT
      integer dimg
      real*8  st,dr,rp
      real*8  rg(-imax:imax)
c OUTPUT 
      real*8  xxo(0:imax),yyo(-imax:imax)
      real*8  phi(0:imax)
c INTERNAL
      real*8  xx(0:imax),yy(-imax:imax)
      real*8  dt,psi
      integer i

c check for rp>=rg(dimg-1)
      if(rp.ge.rg(dimg-1)) then
         write(*,*) 'particle out of right grid margin !!!'
         stop
      endif
c field timestep
      dt=dr
c yy(r)=yyo(r-dt)+st*step[rp,rp+dt]-st*step[-rp,-rp+dt] 
c xx(r)=xxo(r+dt)-st*step[rp-dt,rp] 
      yy(-dimg)=0.d0      
      do i=-dimg+1,-1
         yy(i)=yyo(i-1)
         if(-rp.le.rg(i).and.rg(i).lt.-rp+dt) then
            yy(i)=yy(i)-st
         endif
      enddo
      do i=0,dimg-1
         xx(i)=xxo(i+1)
         yy(i)=yyo(i-1)
         if(rp-dt.le.rg(i).and.rg(i).lt.rp) then
            xx(i)=xx(i)-st
         elseif(rp.le.rg(i).and.rg(i).lt.rp+dt) then
            yy(i)=yy(i)+st
         endif
      enddo
      xx(dimg)=0.d0
      yy(dimg)=yyo(dimg-1)
c rewrite xx and yy
      do i=-dimg,-1
         yyo(i)=yy(i)
      enddo
      do i=0,dimg
         xxo(i)=xx(i)
         yyo(i)=yy(i)
      enddo
c integrate x+y starting from the origin
c using trapezoidal method (order dr**3)
      psi=.5d0*(xx(0)+yy(0))
      do i=1,dimg
         psi=psi+xx(i)+yy(i)
c the gravitational potential is
         phi(i)=dr*(psi-.5d0*(xx(i)+yy(i)))/rg(i)
      enddo
      phi(0)=phi(1)
c check boundary condition at r=0
      if(xx(0).ne.yy(0)) then
         print *,'xx(0) <> yy(0) !!!!!!!!'
      endif

      return
      end
\end{verbatim}
\begin{verbatim}
cccccccccccccccccccccccccccccccccccccccccccccccccccccccccccc
c Given rp (and ur=0) solves for a and up in:
c 1.0)   ut = sqrt(1+ur**2+(rp*up)**2)
c 1.1)   a  =-exp(a/rp)*mr/ut
c 2  )   up = sqrt(-a/(2*rp**3))
c rewritten as
c      -a = exp(a/rp)*mr/sqrt(1-a/(2*rp))
c
c INPUT   mr (= rest mass),rp (= shell radius)
c OUTPUT  a ("potential"), up (= angular velocity)
cccccccccccccccccccccccccccccccccccccccccccccccccccccccccccc
      subroutine phi1(mr,rp,a,up)
      implicit none
c INPUT
      real*8  mr,rp
c OUTPUT
      real*8  a,up
c INTERNAL
      real*8  ao,sqti,ep,f,fp
      real*8  acc
      parameter(acc=1.d-15)

      a=0.d0
c start the Newton iteration
 10   sqti=1.d0/sqrt(1.d0-a/(2.d0*rp))
      ep=exp(a/rp)
      f=a+mr*ep*sqti
      fp=1.d0+mr*ep*(sqti+.25d0*sqti**3)/rp
      ao=a
      a=ao-f/fp
      if(abs(f).gt.acc) goto 10

c the angular velocity rp*up**2=(a/rp**2)/2
      up=sqrt(-a/(2.d0*rp**3))
      return
      end
\end{verbatim}
\begin{verbatim}
cccccccccccccccccccccccccccccccccccccccccccccccccccccccccccc
c        Finds the equilibrium radius
c given utlp solves for a and rp in: 
c 1.0)   utt    = sqrt(exp(2*a/rp)+(utlp/rp)**2)
c 1.1)   a      =-exp(2*a/rp)*mr/utt
c 2)     utlp**2=-exp(2*a/rp)*rp**3*(a/rp**2)/2
c rewritten as
c     y=4*a**4/(a**4-(2*utlp*mr)**2)
c 1) utlp**2*(y/a**2)=-exp(y)         -------> find a (<0)
c 2) r=(a**4-(2*utlp*mr)**2)/(2*a**3) -------> find r (>0)
c INPUT   utlp (= angular momentum), mr (= rest mass)
c OUTPUT  erp (= equil  radius),ea (=  equil "potential")
cccccccccccccccccccccccccccccccccccccccccccccccccccccccccccc
      subroutine eqrp1(utlp,mr,erp,ea)
      implicit none
c INPUT
      real*8 mr,utlp
c OUTPUT
      real*8 ea,erp
c INTERNAL 
      real*8 a4,y,ai,af,afo,fi,ff
      real*8 acc,u2,fu2m2
      parameter(acc=1.d-10)

      u2=utlp**2
      fu2m2=4.d0*u2*mr**2
c upper limit
      ai=0.d0
      fi=1 
c find the lower limit
      af=-sqrt(2.d0*u2*(-1.d0+sqrt(1.d0+(-mr/utlp)**2)))
c start the secant iteration
 10   a4=af**4
      y =4.d0*a4/(a4-fu2m2)
      ff=u2*y/af**2+exp(y)
      afo=af
      af=afo-ff*(afo-ai)/(ff-fi)
      if(abs((af-afo)/afo).gt.acc) then
         ai=afo
         fi=ff
         goto 10
      endif
c found a find r
      erp=(af**4-fu2m2)/(2.d0*af**3)
      ea =af

      return
      end
\end{verbatim}
\begin{verbatim}
cccccccccccccccccccccccccccccccccccccccccccccccccccccccccccccccc
c  writes parameters in output
cccccccccccccccccccccccccccccccccccccccccccccccccccccccccccccccc
      subroutine out(mr,ri,rf,xi,dt,dr,dimt,dimg)
      implicit none
      real *8 mr,ri,rf,dt,dr,xi
      integer dimt,dimg

c rest mass
      write(*,*) 'mr=',mr
c initial radius
      write(*,*) 'ri=',ri
c xi=up/up(circular)
      write(*,*) 'xi=',xi
c final equilibrium radius
      write(*,*) 'rf=',rf
c time step
      write(*,*) 'dtf=',dt
c grid spacing
      write(*,*) 'dr=',dr
c total integration time=dimt*dtf
      write(*,*) 'dimt=',dimt
c grid dimension r in [0,dimg*dr]
      write(*,*) 'dimg=',dimg
      
      return
      end
\end{verbatim}
}
\newpage \noindent
This is the code used for enveloping the numerical integration.
{\tiny
\begin{verbatim}
ccccccccccccccccccccccccccccccccccccccccccccccccccccccccccccccc
c  Relaxation to virial equilibrium
c
c INPUT  r0=shell initial radius
c        mr=shell rest mass
c        xi=up/up(circular)
c
c OUTPUT fort.10 : r_max,utt(r_max),xi
c        fort.14 : time,r_max
ccccccccccccccccccccccccccccccccccccccccccccccccccccccccccccccc
      implicit none
c INPUT
      real*8  r0,mr,xi
c OUTPUT
      real*8  utt,r(0:1000),tt
c INTERNAL
      real*8  dr,dtdr(0:1000),etot,dedt,dedr
      real*8  a,ea,rf,utlp,up,am2
      real*8  e2p,npo2pi,xi2
      integer i,np
      parameter(np=999)
c ========================INPUT DATA============================
      write(*,*) 'initial radius r0'
      read(*,*) r0
      write(*,*) 'rest mass mr'
      read(*,*) mr
      write(*,*) 'ratio utlp/utlp(circular)'
      read(*,*) xi 
      if(xi.ge.sqrt(2.d0)) then
         print *,'qust.f uses Newtonian approx. : xi < sqrt(2)'
         stop
      endif
c ======================INITIALIZATION==========================
c find angular velocity for the circular orbit at r0
      call phi1(mr,r0,a,up)
      utlp =xi*r0**2*exp(a/r0)*up
      am2  =utlp**2
c find final equilibrium radius rf and particle energy
      call eqrp1(utlp,mr,rf,ea)
      dr   =(rf-r0)/dble(np)
      print *,'initial radius, potential=',r0,a/r0
      print *,'final   radius, potential=',rf,ea/rf
      print *,'initial energy utt=',sqrt(exp(2.d0*a/r0) +am2/r0**2)
      print *,'final   energy utt=',sqrt(exp(2.d0*ea/rf)+am2/rf**2)
c =================r_{max},xi,utt(r_{max})======================
      do i=0,np-1
         r(i)=r0+dble(i)*dr
         call phi1(mr,r(i),a,up)
         e2p=exp(2.d0*a/r(i))
         utt=sqrt(e2p+am2/r(i)**2)
c the new xi at r(i) is
         xi=utlp/(exp(a/r(i))*up*r(i)**2)
c write fort.10 : (r,utt,xi)
         write(10,*) r(i),utt,xi
c the total energy is then
         etot=mr*utt
         xi2=xi**2
c calculate detot/dr
         dedr=(a*r(i)*(4.d0-7.d0*xi2)+a**2*2.d0*xi2+r(i)**2*4.d0
     $  *(xi2-2.d0))/(xi2*r(i)**3*(7.d0*a*r(i)-2.d0*a**2-
     $   4.d0*r(i)**2))
         dedr=mr*dedr*am2/utt
c calculate detot/dt
         npo2pi=sqrt(2.d0*r(i)**3/(a*(xi2-2.d0)**3))
         dedt=-(sqrt(2.d0)/9.d0)*mr**2*(sqrt(-a)**5/sqrt(r(i))**7)
     $        *((1.d0-xi2)**2/xi**7)*(5.d0-2.d0*xi2+xi**4)/npo2pi        
c calculate dr/dt
         dtdr(i)=dedr/dedt
      enddo
c ====================t,r_max===================================
      write(14,*) 0,r0
c integrate (dt/dr) to get t(r) 
      tt=.5d0*dtdr(1)
      do i=1,np-1
         tt=tt+dtdr(i)
c make graph (t(r),r)
         write(14,*) dr*(tt-.5d0*dtdr(i)),r(i)
      enddo

 30   stop
      end
\end{verbatim}
}
\newpage \noindent
This is the code used for the quasistatic integration.
{\tiny
\begin{verbatim}
cccccccccccccccccccccccccccccccccccccccccccccccccccccccccccccccc
c            1 SHELL QUASISTATIC CLUSTER                           
c
c INPUT rp=shell radius
c       mr=shell rest mass
c       xi=up/up(circular)
c       dt=integration timestep
c       time=simulation duration      
c OUTPUT              
c       fort.18 = (t,rp)
c       fort.19 = (rp,utt,utlr)
c       erp=equilibrium radius
cccccccccccccccccccccccccccccccccccccccccccccccccccccccccccccccc
      implicit none
c INPUT
      real*8  rp,mr,xi,dt,time
c OUTPUT
      real*8  utlr,utt,erp,ea
c INTERNAL
      real*8  am2,e2p,e2ppr
      real*8  up,a,utlp
      integer tsteps,dimt
c ========================INPUT DATA============================
      write(*,*) 'initial radius rp'
      read(*,*) rp
      write(*,*) 'rest mass mr'
      read(*,*) mr
      write(*,*) 'ratio utlp/utlp(circular)'
      read(*,*) xi 
      write(*,*) 'time step dt'
      read(*,*) dt
      write(*,*) 'time lenght'
      read(*,*) time
      dimt=int(time/dt)
c =====================INITIAL CONDITION========================
      tsteps=0
c ----------particle------------------------------------
c tangential orbit (utlr=0)
      utlr=0.d0
c find angular velocity for the circular orbit at rp
      call phi1(mr,rp,a,up)
c angular momentum for circular orbits (in a time
c independent field) is:
c utlp(circ)=ulp(circ)*exp(phi)=up(circ)*r*r*exp(phi)      
c set utlp=xi*utlp(circ) = constant of motion
      utlp=xi*rp**2*exp(a/rp)*up
      am2 =utlp**2
c ----------field---------------------------------------
c phi(r)=a/rp for r<=rp
c phi(r)=a/r  for r> rp
c =====================NEXT TIMESTEP============================
 100  tsteps= tsteps+1
      e2p=exp(2.d0*a/rp)
c find e2ppr=e2p*(phi,r(rp-)+phi,r(rp+))/2
      e2ppr=-e2p*.5d0*a/rp**2
c ----------particle-----------------------------------------
c evolve with 4-th order Runge-Kutta         
      call runge4(am2,e2p,e2ppr,dt,rp,utlr,rp,utlr) 
      if (rp.le.0.d0) then
         print *,'particle fallen in to the origin !!!'
         stop
      endif
c ----------field--------------------------------------------      
      call phi1(mr,rp,a,up)

c write fort.18 :[t,r(t)] and fort.19 :[t,utt(t),utlr(t)]
      write(18,*) tsteps*dt,rp,a/rp
c calculate utt
      utt=sqrt(e2p+utlr**2+am2/rp**2)
      write(19,*) rp,utt,utlr
      if(tsteps.eq.dimt) goto 200
      goto 100

c estimate the final equilibrium radius erp
 200  call eqrp1(utlp,mr,erp,ea)
c write erp and the field at erp (ea/erp)
      print*,erp,ea/erp

      stop
      end
\end{verbatim}
}
\onecolumn

\listoffigures

\bibliographystyle{unsrturl}
\bibliography{bibrelax}

\begin{thebibliography}{1}

\bibitem{Shapiro92}
{S. L. Shapiro and S. A. Teukolsky}.
\newblock {Scalar gravitation: A laboratory for numerical relativity}.
\newblock {\em Phys. Rev. D}, 47:1529, 1992.
\newblock \href {https://doi.org/10.1103/PhysRevD.47.1529}
  {\path{doi:10.1103/PhysRevD.47.1529}}.

\bibitem{Godunov59}
S.~K. Godunov.
\newblock {A Finite Difference Method for the Numerical Computation of
  Discontinuous Solutions of the Equations of Fluid Dynamics}.
\newblock {\em Mat. Sb.}, 47:271, 1959.

\end{thebibliography}

\end{document}